\documentstyle[12pt,aasms4,amssym]{article}

\lefthead{Allende Prieto, Garc\'{\i}a L\'opez, Lambert \& Gustafsson}
\righthead{Spectroscopic gravities as compared with {\it Hipparcos} data}
\begin{document}

\title{A consistency test of spectroscopic gravities for late-type stars}

\author{Carlos Allende Prieto\footnote{Present address: McDonald Observatory 
and Department of Astronomy, The University of Texas at Austin, RLM 15.308, Austin, TX 78712-1083, USA} and Ram\'on J. Garc\'{\i}a L\'opez}
\affil{Instituto de Astrof\'\i sica de Canarias \\ E-38200, La Laguna,
Tenerife,  \\  SPAIN}

\author{David L. Lambert}
\affil{McDonald Observatory and Department of Astronomy, \\ The University of Texas 
at Austin \\ RLM 15.308, Austin, TX 78712-1083, \\ USA}

\author{Bengt Gustafsson}
\affil{Uppsala Astronomical Observatory \\ Box 515, S-751 20 
Uppsala \\ SWEDEN}

\authoraddr{IAC, E-38200, La Laguna, Tenerife,  SPAIN}

\authoremail{callende@iac.es, rgl@iac.es, dll@astro.as.utexas.edu 
and bg@astro.uu.se}

\begin{abstract}

Chemical analyses of late-type stars are usually carried  out  following the
classical recipe: LTE line formation and homogeneous, plane-parallel,
flux-constant,  and LTE model atmospheres.  We review different results 
in the literature that have suggested  significant  inconsistencies in the
spectroscopic analyses, pointing out the difficulties in deriving independent
estimates of the stellar fundamental parameters and hence,
detecting systematic errors. 

The  trigonometric parallaxes measured by the {\it Hipparcos} mission
provide accurate appraisals of the stellar surface gravity for nearby
stars, which are used here to check the gravities obtained from the
photospheric iron ionization balance. We find an approximate agreement
  for stars  in the metallicity range $-1.0 \le$ [Fe/H] $\le 0.$, but the
comparison shows that the differences between the spectroscopic and 
trigonometric gravities decrease towards lower metallicities
 for more metal-deficient dwarfs ($-2.5 \le$ [Fe/H] $\le-1.0$), which
 casts a shadow upon the abundance analyses for   extreme metal-poor stars
that make use of the ionization equilibrium to constrain the gravity.
The comparison with  the strong-line gravities derived by Edvardsson (1988)
and Fuhrmann (1998a) confirms  that this method provide systematically
larger gravities than the ionization balance. The strong-line gravities get
closer to the physical ones for the stars analyzed by Fuhrmann,
but they are even further away than the iron ionization gravities for the stars 
of lower gravities  in Edvardsson's sample. 

The confrontation of the deviations of the iron ionization gravities 
in metal-poor stars reported here  with  departures  from 
the excitation balance  found in the literature, show that they are likely to
be induced  by the same physical mechanism.
\end{abstract}

{\it Subject headings}: radiative transfer -- techniques: spectroscopic --  
stars: atmospheres -- stars: distances -- stars: fundamental parameters --
stars: late-type

\section{Introduction}
 
Interpretation of stellar spectra to quantify  chemical abundances
requires, among other ingredients, the use of model  atmospheres,  
and  our knowledge of the
physics they are built on will  determine  the  accuracy  of  the
analyses.  Following the standard method, measurements   
of equivalent widths or   line profiles at high resolution 
and high signal-to-noise ratio
are analyzed using a classical model atmosphere based on the  following
assumptions:   local thermodynamic equilibrium (LTE),  hydrostatic
equilibrium, conservation of flux, and plane-parallel
 stratification. Also, the   mixing-length  theory is used to take
convection into account. Finally, a  chemical composition is assumed, 
and the non-thermal broadening of spectral lines is treated by a
simplified recipe: micro and macro turbulence (for a review see, e.g.
Gustafsson \& J$\o$rgensen 1994).

Such standard analyses may be subjected to external and internal
consistency checks. Often, such a check involves a determination of a
stellar parameter by two or more methods. An inconsistency may indicate
either an inadequate implementation of the methods for deriving the
stellar parameters or a failure of one or more of the
 assumptions on which the models are built. Of course, both
contributors of inconsistencies may operate simultaneously, and in some
cases, the distinction between them may be subtle. If
the implementation of the separate methods is accurate, consistent
results for a stellar parameter  is
a necessary but not a sufficient indication that a given
 model atmosphere is a reliable representation of the real atmosphere.

Important failures  to reproduce observations have been claimed, even
for the Sun and  other  solar-like  stars.  The H$\alpha$ and
H$\beta$ wings of the solar spectrum cannot be reproduced by some
theoretical models using a typical value for the mixing-length
parameter $\alpha \sim$ 1.5, as  noticed by Fuhrmann,
 Axer \& Gehren  (1993) and Van't Veer-Menneret \& Megessier (1996).
They suggest a lower value for $\alpha$, but this has been rejected by
Castelli, Gratton \& Kurucz (1997),  who showed that the predicted flux
distribution is not in agreement with  observations.
This failure might just be classified as an inadequate implementation
of the classical method.

Tests on the consistency of surface gravities derived by different
procedures have been reported in the literature.  Brief remarks on three
 such  tests must suffice here. Edvardsson (1988) discovered from the
 analysis of a sample of eight sub-giants that the ionization
equilibrium of iron and silicon, as derived from weak neutral and
ionized lines, gave gravities systematically lower  than those obtained
from the pressure-broadened wings of strong metal lines.  Even more
striking was the discrepancy presented by  Fuhrmann et al.  (1997; also
Steffen 1985) for Procyon.  They found $\log g$\  = 3.5  (in c.g.s.
units) from the ionization equilibrium of iron, in strong contrast to
the much higher values they derived from the wings of strong lines and
the astrometric measurements: log $g$ $\simeq$ 4.05.  Gratton, Carreta
\& Castelli (1996) derived gravities for giants in globular clusters
using theoretical isochrones, and compared them with the spectroscopic
gravities (requiring ionization equilibrium), detecting large
differences for  lower effective temperatures.  The
  results of Nissen, H$\o$g \& Schuster (1997) suggest that the
spectroscopically derived gravities  in the
 literature are systematically lower than those derived from {\it
Hipparcos} (ESA 1997) parallaxes for the metal-poor stars in the
gravity range $3.5 \le \log g \le 4.5$.

Very recently, Fuhrmann (1998a) has carried out a similar check of the
spectroscopically derived gravities, using them to estimate {\it
spectroscopic} distances which he compares with the {\it Hipparcos}
data for nearby (i.e. with highly accurate parallaxes) stars. This kind
of comparison  is equivalent to that of  Nissen et al. (1997)
 but using  distances instead of gravities.  His analysis covered stars
 in the metallicity range\footnote{[M/H] = $\log (\frac{N(M)}{N(H)}) -
\log (\frac{N(M)}{N(H)})_{\odot}$  where N(M) is the number density of
the nuclei of the element M and "H" refers to hydrogen.}  $-1.95 \le$
[Fe/H] $\le +0.40 $ and gravities running the gamut from $\log g = 3.1$
to 4.7. He finds an impressive agreement between distance
estimates, as indicated by the mean difference which is  a mere 3.4 \%,
with a scatter of 4.9 \% (after rejecting a few outliers). This result
is not in contradiction with the differences found by Nissen et al.
(1997), as the spectroscopic gravities of Fuhrmann were not derived
from the ionization equilibrium only, but  a combination of this
estimator with the pressure sensitive wings of strong singly ionized
metal lines. The discrepancy  between the two methods was confirmed
again for  the hotter stars, for which the strong-line gravities were
preferred.

Another series of consistency checks involves the stellar effective
temperatures ($T_{\rm eff}$). Observationally, the checks exploit
either the flux distribution or the excitation of a species (Fe~I, for
example) well represented by weak lines in the stellar spectrum. Again,
a few examples must suffice here.  Magain  (1985)  found discrepancies
between the ultraviolet  fluxes predicted by the models and the
observations for two metal-poor stars (see also Gustafsson \& Bell 1979),
 although Bell \& Oke (1986)
arrived at the opposite conclusion. Magain \& Zhao (1996)   reported
strong correlations  between  iron abundances and the excitation
potential of the employed spectral lines of neutral iron, i.e., the
Fe~I excitation temperature is not that predicted by the classical
model of the photometry-based (b-y and V-K calibrations from
Magain 1987) effective temperature. They
quantified the departures from the excitation equilibrium and mapped 
 them on the $T_{\rm eff}-\log  g$ plane, 
finding smooth variations of the slope
of the derived iron abundance against the excitation potential. However, we
note that the metal-poor sub-giant HD140283, as analyzed 
by Magain (1984), exhibits a slope of the iron
 abundance with  excitation potential in the opposite sense to that
expected from the position of the star in the  $T_{\rm eff}-\log  g$ diagram
and the picture drawn by Magain \& Zhao (1996).  Analysis of the
excitation balance in the metal-poor giant HD122563 by Dalle Ore (1993)
showed that departures were present, and Gratton et al. (1996) found
that the departures do not depend on gravity.

However,  conclusions  must be extracted with extreme care. 
The situation is especially complex, as many factors  influence the
consistency tests. Errors in the photometric effective temperatures,
inadequacies in the damping treatment, or systematic errors in the
atomic data or the spectroscopic measurements are enough to produce an
apparent disagreement, which might be wrongly interpreted as a fault of
the model atmospheres.

Devotees of metal-poor stars wonder if the suggested inconsistencies,
often demonstrated for more metal-rich stars too,
 are  more or  less severe for their favorite stars that hold the keys
to the chemical evolution of the early Galaxy.  As pointed out by
Gustafsson \&  J$\o$rgensen (1994), or Magain \& Zhao (1996), two
points  seem  to  be especially  critical  for   the   low metallicity
stars:

      a) LTE:  due to the  low  metallicity and the therefore much weaker 
metal-absorption in the ultraviolet, a more non-local UV flux is able to 
penetrate from the deeper layers. This flux is vital in determining 
the ionization equilibrium of the atoms. As a consequence, the role of radiation 
on the thermodynamical state of matter becomes more important, resulting in 
stronger deviations from LTE.

   b)  Inhomogeneities:  the  lower  H$^{-}$ due to diminished amount of
electrons makes the atmosphere more transparent so that the effects
of convection can be more clearly seen in the  photosphere. This is
 suggested by the results  of Allende Prieto et al. (1995, 1999), who found
 convective line asymmetries to be stronger than expected 
in the metal-poor  subgiant HD140283.

These two points could be even more critical for the 
low-gravity stars where, due to the low densities, both deviations 
from LTE and  convective temperature inhomogeneities may be enhanced.

In this work, we have extended the analysis of trigonometric gravities
 to many stars for which spectroscopic gravities had been previously 
estimated from the iron ionization equilibrium. Assessment of the 
spectroscopic gravity scale should be considered a mandatory step, to
properly connect   chemical abundances of stars in globular clusters, which
are derived from the spectra of red giants,  
with those of fields halo stars, which  come from dwarfs or slightly
 evolved  subgiants.

\section{`Trigonometric' Gravities}

The newly available {\it Hipparcos} parallaxes  have provided distances
to many stars in the solar neighborhood. These  measurements allow us
to constrain the stellar gravities. In this section, the
trigonometric gravities are compared with the gravities  obtained
spectroscopically from  the neutral and ionized iron lines.  Our goal
is to verify  the results of Nissen et al. (1997),
 extending their study to many more stars, and to determine if the
differences between trigonometric and spectroscopic gravities
 depend primarily on gravity, metal content,  temperature, or a
combination of all three parameters.

The  scheme uses the familiar
relationships: $g \propto M/R^2$ and  $L \propto R^2T^4_{\rm eff}$, where 
$M$, $R$, and $L$ are the stellar mass, radius, and luminosity,  respectively.
Straightforward manipulation gives

\begin{equation}
\label{logg}
{\rm log} \frac {g} {g_{\odot}} = {\rm log} \frac {M} {M_{\odot}}
  + 4{\rm log} \frac {T_{\rm eff}} {T_{{\rm eff},\odot}} +0.4 V_0
  +0.4{\rm BC} +2{\rm log}\pi +0.12,
\end{equation}

\noindent where the $\pi$ is the parallax,  $V_0$ the 
 apparent Johnson $V$ magnitude, and $BC$ the bolometric correction.
 Nissen et al. (1997)  have already applied {\it Hipparcos} parallaxes
 in this way -- see also Fuhrmann (1998b) -- 
and we  have followed their recipe. Alternatively, 
Equation 1 may be used to derive  a spectroscopic parallax  
from a spectroscopic gravity,  as done by Fuhrmann (1998a). 
We have assumed that the reddening is negligible, 
as the stars are all nearby. A theoretical 
isochrone  is used to estimate the stellar mass and the bolometric
 correction (BC) from the absolute $M_v$ 
magnitude\footnote{No extrapolation  was accepted here, 
and a star was excluded from the study when its mass 
and BC could not be obtained by interpolation in the theoretical isochrone 
of the  metallicity assigned.},  and they are combined with the {\it
Hipparcos} parallax and the effective temperature 
to get the stellar gravity. The derived gravities
are not very sensitive to  errors in the  assumed parameters. Thus,
an error of  3\% in  $T_{\rm eff}$  typically produces
uncertainties in the derived $\log g$\  of 0.05 dex, a 30 \% error in the mass
translates to 0.13 dex, and an error of 30 \% in the bolometric
correction to a 0.05 dex error in the $\log g$. Nissen et al. (1997) have 
already argued that statistical biases in the parallaxes 
in a sample of this kind are unlikely to occur.  
We   comment more on   errors below.

 We have  collected data for  F-K stars nearby stars (see below)
from  Bonnell \& Bell (1993):  4 stars; Edvardsson (1988): 7  stars; 
Fuhrmann et al. (1997): 9 stars; Gratton et al. (1996): 214
 stars; Magain (1989): 8 stars; and Ryan, Norris \& Beers (1996): 1
 star. In some of these works the spectroscopic gravities were derived
from both the ionization balance and the wings of strong lines. These
two criteria are known to differ (Edvardsson 1988, Fuhrmann 1998a) and
we have first restricted the comparison to the ionization equilibrium
gravities, which has been the procedure most extensively used.
  Bonnell \& Bell (1993) derived gravities from the 
comparison of atomic and molecular magnesium features, but they also
estimate and compile iron ionization balance gravities from the literature, 
which we discuss here.

Most of the data come from Gratton et al.  (1996) who reanalyzed 
in a consistent
fashion stars for which adequate and reliable data exist in the
literature.  Stellar effective temperatures were estimated afresh by
Gratton et al.  using the Infrared Flux Method (Blackwell  \&
Lynas-Gray 1994) while the gravities were obtained from the ionization
equilibrium of iron. For their analysis, Gratton et al. use the Kurucz
(1992) model atmospheres.  The combined sample totals more than 300
stars with metallicities from [Fe/H] $\sim -3$ to $\sim$ 0.2.  We shall
not go deeper into details on the different effective temperature
scales, model atmospheres, etc., employed by the different sources at this
point.  Several stars are common to two or more of
these references and, in some cases, the discrepancies are large.

 We have taken  the published effective temperatures 
and  derived the trigonometric gravities
for those stars for which the {\it Hipparcos} parallax exceeds 10
mas\footnote{miliarcseconds.}.  Perryman et al. (1995) showed that  the
error of the parallax is typically 
less than 20\% in such cases.  Oxygen-enhanced
isochrones  by  Bergbusch \& Vandenberg (1992) were used within the 
published range, i.e. $-2.26 \le $ [Fe/H] $\le -0.47$.  For the 
stars falling outside this range, the appropriate extreme value was
used.   An age of 8 Gyr. was assumed for stars  with  [Fe/H] $> -0.47$.  
 while an age of 12 Gyr. was adopted for those  with [Fe/H] $\leq -0.47$.
In principle, an individual determination of age and mass is 
possible  with the present data, but the more schematic approach 
chosen here was found to be adequate because  reducing the
age assumed for the metal-poor stars from 12 Gyr. to 8 Gyr.
 increases the derived gravities by less than 0.06 dex for most of the stars.
Alternative and likely improved isochrones have been published, but
adopting those would have little effect on the derived 
trigonometric gravities, as we discuss below. In
particular, newer calculations use the improved OPAL opacities
(Iglesias et al. 1992), as those, e.g., by Bertelli et al. (1994).
We find that use of Bertelli's isochrones, which  do 
not take  $\alpha$-elements enhancement for 
metal-poor stars into account,  results in very small changes to the
derived gravities, always below 0.1 dex.  

The distribution of
relative errors in the parallaxes of the sample, $\frac{\sigma}{\pi}$, is shown
in Fig. 1.  Most stars have relative errors in the
parallax below  5\%.   Supposing that  errors in the different
 quantities  included in Equation \ref{logg} are  independent from each
 other, we find 

\begin{equation}
\label{error}
\begin{array}{ll}
\sigma^2(\log g) = \log^2 e & 
\left[\left(\frac{\sigma(M/M_{\odot})}{M/M_{\odot}} \right)^2 + 
16 \left(\frac{\sigma(T_{\rm eff})}{T_{\rm eff}} \right)^2 + 
 4 \left(\frac{\sigma(\pi)}{\pi} \right)^2 + \right. \\
&\left. 0.16  \left(\frac{\sigma(F_V)}{F_V}\right)^2 + 
0.16 \left(\frac{\sigma(F_{BC})}{F_{BC}}\right)^2 \right] 
\end{array}
\end{equation}

\noindent where $F_V$ and $F_{BC}$ are the fluxes in and out the V band, 
respectively, and then, assuming  errors in the stellar mass are of the 
order of 10\%, the typical uncertainties in  $T_{\rm eff}$\   
of $\sim$ 3\%, those
in the V photometric band and the bolometric corrections of about 5\%,
 the derived $\log g$\  will have an uncertainty of 0.08 dex. For the
 case  $\frac{\sigma}{\pi} \simeq$ 0.08, the uncertainty of the
$\log g$\  will be $\simeq$ 0.10 dex.

The included stars, listed in Table 1,  are shown in Figure 2 where we
also plot representative theoretical isochrones.  Metal-poor stars are
shown by the filled circles ([Fe/H] $\le$ --0.47), while the open
circles correspond to those richer than [Fe/H] = --0.47.  Some stars
have multiple entries in Table 1 and all figures, as they have been analyzed in
more than one of the included references.  There are also some stars
that have multiple entries in the work of Gratton et al. (1996), 
corresponding to the different sources  they took the equivalent widths
from.  Nissen et al.  (1997) found that a  
shift of 100 K in the effective temperature
of the isochrones was needed to match their sample of stars to the
isochrones,  and pointed out that the shift is equivalent to a decrease
of the convection parameter $\alpha$ to a value  less than unity.  From
the inspection of Fig. 2, we cannot conclude whether a shift is needed or not
for the present sample. (The temperature scale of Gratton et al. 1996
has been used for most of the stars and in fact, Gratton et al. have
shown that there are important systematic
 discrepancies between their $T_{\rm eff}$ scale
and that of Edvardsson et al. 1993 and Nissen et al.  1994).  A
detailed comparison shows that there are systematic and much more
complicated deviations than a constant shift between the $T_{\rm eff}$s 
used here (see Table 1) and those included in the isochrones.

Results from the compiled sample are first shown in Figure
3 where we plot $\log g$\  from spectroscopy versus $\log g$\ 
from the trigonometric parallaxes for a metal-poor and a metal-rich
sample, where [Fe/H] = --0.47 is again taken as the boundary 
between both.  At first glance, there are not large
differences between the spectroscopic (iron ionization)
  and trigonometric gravities:  the mean difference between the two 
gravity estimates is $\Delta \log g = 0.08 \pm 0.21$ 
(1$\sigma$) dex. Although this value has been drawn 
combining data from different sources, the comparison 
source by source 
reflects that the scatter is not greatly reduced (see Table 2).  
If the spectroscopic gravities are correct to
0.15 dex, and the trigonometric estimates 
have errors of 0.08 dex (as previously estimated for most of the stars studied
here), the rms difference should be about 0.17 dex, 
in agreement with the results for the individual
works in Table 2  (0.15 dex--0.25 dex). Systematic deviations between the 
spectroscopic gravities  derived from different groups arising from the 
use of different  effective temperature scales, atomic data, 
model atmospheres, etc. account well for  the rms difference
 for the compiled sample (0.21 dex). It is apparent from Fig. 3 that the
metal-poor stars show the largest deviations, and a clear correlation can
be see in Table 2 between the mean metallicity of the subsamples and the
mean of  deviations of the spectroscopic gravities from the trigonometric
values. This is studied in detail  in the next section.

\section{The effect of the metallicity on the spectroscopic gravities.}

The overall agreement, within the expected errors, between the
spectroscopic and the trigonometric gravities found in the previous section
can be analyzed in more detail by looking for dependences of the 
discrepancies on the stellar properties. The different works combined
in this analysis correspond in some cases to small (this does not apply to
Gratton et al. 1996, which includes more than 200 nearby stars), 
 homogeneous  samples of  stars, such as the metal-rich  giants 
in Bonnell \& Bell (1993) and Edvardsson (1988),  or the metal-poor 
stars analyzed by  Magain (1989) and Fuhrmann et al. (1997). Combining
all these studies, we provide a  general perspective by covering
a larger zone in the space of the stellar parameters. Nevertheless, special
care has to be taken, as systematic effects may be present among 
 the different subsamples.

Figure 4a  shows the differences found in Section 2 as a function of
the stars' effective temperature and gravity, while Fig. 4b displays the 
same differences against metallicity. 
The different subsamples are identified with distinct
symbols. While no correlation is present between the discrepancies and
the gravity or the effective temperature, a trend  with metallicity is
apparent. However, metal-poor stars belong mainly to the galactic halo,
in opposition to more metal-rich stars which correspond to the disk
--in a very simplified picture--, and hence they tend to be more rarely
present in the solar vicinity (not very far from the galactic plane)
than solar-metallicity stars. This makes the metal-poor stars in the
sample tend to be at larger distances and therefore, the relative
errors in their parallaxes are larger than those for metal-rich stars.
Fig. 5 (top panel) shows this effect. Restricting the analysis to the
 stars with uncertainties in the parallax between 5 \% and 10 \%
 eliminates  the systematic biases, as confirmed by the histogram in
the middle panel of Fig. 5, but the trend of $\log g {\rm (Spec.)} -
\log g {\rm (Trig.)}$ with the metallicity  seen in Fig.  4b  persists
(see Fig. 5, lower panel).

The question of whether systematic effects in the trigonometric
gravities may arise from the stellar evolutionary models and the
assumed stellar ages can be addressed  by repeating the calculations
using a different set of models. The upper panel in Fig. 6 shows the
differences between the gravities derived using the isochrones of
Bertelli et al. (1994)  and those of Bergbusch \& Vandenberg (1992). Again,
the nearest  extreme metallicity  in the published grid  was employed for the
stars outside the metallicity range of the models ($-1.7  <$ [Fe/H]
$< 0.4$ for Bertelli et al). The lower panel reflects the differences 
in the gravities when assuming the same metallicity for all  stars 
([Fe/H]=--0.47) and  the same age (12 Gyrs.), making use of Bergbusch \&
Vandenberg's isochrones. These differences cannot
 account for the systematic effects shown in Fig. 4b for  stars more  
metal-poor  than [Fe/H]=--1.0. 

\subsection{[Fe/H] $> -1$}

This sample of stars  is strongly dominated by the
data from Gratton et al. (1996). The stars analyzed by 
Bonnell \& Bell (1993) and Edvardsson (1988) have been introduced
 here to expand the sample of stars with large
gravities (see Fig. 4a, upper panel).

The mean difference between the spectroscopic and the trigonometric
gravities is $0.10 \pm 0.17 $ (1$\sigma$) dex, and a linear regression
analysis confirms the visual impression from Fig. 7a in the sense that  none of
the subsamples exhibits a statistically  significant slope of the
gravity discrepancy with metallicity. We note that the rms deviation is
in agreement with our expectations, as described in \S2, assuming the
spectroscopic values have errors of 0.15 dex.

\subsection{[Fe/H] $\le -1$}

Fig. 7b differs from an expansion of Fig. 4b  in
that the stars of Magain (1989) have been shifted in metallicity by $+0.17$
dex to take into account that his analysis was not differential to the Sun and,
in contrast to Fuhrmann et al.  (1997) and Gratton et al. (1996), he assumed
the solar iron abundance to be significantly larger than the meteoritic value,
 in particular $\log$ N(Fe)$_{\odot} + 12 = 7.68$.  

A clear zero-point shift exists between the spectroscopic (iron
ionization balance) gravities of Gratton et al.  and those derived by
Fuhrmann et al. and Magain. Magain and Gratton et al. use laboratory
transition probabilities whenever possible, supplemented with {\it
solar} empirical data obtained with the help of the Holweger \& 
M\"uller (1974) model photosphere. Despite the use of different model
atmospheres, the systematic discrepancy between them can largely 
be ascribed to their effective temperature scales: the  $T_{\rm eff}$s 
assigned by Magain for the 16 stars in common are 
cooler by $-136 \pm 24$  K  than
those of Gratton et al.  Ionization balance establishes that $\Delta
\log g \simeq 0.002 \times \Delta T_{\rm eff}$  (see, e.g., Allende
Prieto 1998, Fig. 1.2), or $\Delta \log g \simeq 0.005 \times 
\Delta T_{\rm eff}$ (following the considerations in \S5) thus 
the offset in the temperature scale accounts for
the  displacement observed in the spectroscopic gravities.  The same
explanation does not apply to the discrepancy  between Gratton
et al. and Fuhrmann, already discussed by Fuhrmann (1998a).  Their
$T_{\rm eff}$ scales, based on the wings of the Balmer lines (Fuhrmann)
and the Infra-Red Flux Method (Gratton et al.), are consistent for
metal-poor stars (Gratton et al. 1996,  Fuhrmann 1998a).  In particular,
the 4 stars in common included in Table 1 show a mean difference of $8
\pm 50 $ K. Several elements in the analyses are different and may be
acting to induce the divergence between the gravity scales, 
 placing Fuhrmann's results closer to those of Magain:  model atmospheres, 
transition probabilities, and equivalent widths.
 
Setting aside  systematic shifts in the gravity scales, our three
sources show the spectroscopic gravities to become smaller relative to
the trigonometric estimates towards lower metallicities. The slopes of
a linear model ($\Delta \log g = a  \times$ [Fe/H] $+ b$), 
obtained by minimizing the $\chi^2$ statistics, are
consistent for the three references: $a = 0.33 \pm 0.08$ (Gratton et al.),
$0.33 \pm 0.06$ (Fuhrmann), and $0.41 \pm 0.16$ (Magain). The
spectroscopic gravity found by Ryan et al. (1996) for HD140283 lies
in between those found by Fuhrmann (or Magain) and Gratton et al.  The
positive slope of the gravity discrepancies  raises the
question of whether this trend will continue towards lower values,
producing important (metallicity dependent) errors in the spectroscopic
gravities for the stars in the domain  [Fe/H] $\le -3$. This would
affect to  the conclusions extracted from 
analyses of chemical abundances in extremely metal-poor stars that
make use of the ionization equilibrium to derive the surface gravity.
Unfortunately, such halo stars are too distant for an accurate
determination of their parallaxes.  

To make sure that multivariate correlations  are not masking real 
dependencies of the differences between spectroscopic and trigonometric
gravities for metal-poor stars, we have corrected the trend with metallicity
found for  the sample of Gratton et al. and check for correlations with the
other parameters, namely $\log g$ and $T_{\rm eff}$. The result is again
negative, as no trend is apparent.

Other consistency checks of the spectroscopic gravities have been 
performed for   giants in globular clusters such 
as M15 (Sneden et al. 1997) and  NGC7006 (Kraft et al. 1998), 
using the absolute magnitude of RR Lyr stars to estimate the distance
to the cluster. The tests showed agreement within the expected uncertainties:
$\sigma \simeq 0.15$ dex demonstrating that well evolved stars 
of moderately low  metallicity, which are not represented in our sample,  
fit in the   general picture of Fig. 4. Recently, evidence has accumulated for
a brighter luminosity scale for the globular clusters (see, e.g., Reid 1997;
Kov\'acs \& Walker 1999). This would make the physical gravities to be up
to 0.2 dex lower than previously though, but still consistent with Fig. 4b.

\section{Ionization vs. strong-line gravities.}

The sensitivity of the wings of strong neutral metal lines to  atmospheric
pressure has been been employed by many analysts to estimate surface
gravities, as an alternative to the ionization balance. Among the 
spectroscopic  studies compiled here, those of Edvardsson (1988) and 
Fuhrmann et al. (1997)\footnote{Updated from  Fuhrmann (1998a)  
for the stars analyzed here.} make use of this approach. 
Fuhrmann (1998a) has found a good agreement between 
the ionization balance and the strong-line
gravities for stars of about the effective temperature of the Sun, but he
argues in favor of the strong-line values for  hotter stars, as the ionization
equilibrium might be affected from NLTE effects and he indeed detects
significant differences between the two methods for the hotter stars in
his sample. The eight stars with $T_{\rm eff} > 5900 $ K 
whose gravities were determined using the ionization 
equilibrium (Fuhrmann et al. 1997), and
 from the analysis of Mg I strong lines by
Fuhrmann (1998a) show that the strong-line gravities are  preferred. 
This is shown in Fig. 8. The ionization balance 
gravities (filled circles)  depart from the 
trigonometric values in $-0.20 \pm 0.12$ (1$\sigma$) dex, 
while the strong-line  gravities (open circles) get closer, 
dropping the difference down to  $-0.01 \pm 0.13$ dex. 
Regrettably, Edvardsson's analysis of metal-rich evolved stars 
shows a different picture. Ionization equilibrium gravities for the
 seven stars compared here show a mean difference with the trigonometric
gravities of $+0.19 \pm 0.15$ dex, while the strong-line method leads to
larger departures, providing a mean of $+0.50 \pm 0.17$ dex. 

The discrepant results may arise from  different elements
 in the analysis (model atmospheres, atomic data, damping treatment,
 etc.) or may emerge naturally from differences in temperature and
gravity covered by the samples. Edvardsson used the $T_{\rm eff}$ 
scale of Frisk (1983), which was built upon flux-constant model 
photospheres. There is no simple way of making a 
meaningful comparison between Frisk's and  Fuhrmann's  
temperature scales. We can only  conclude that
strong-line gravities are systematically larger than those derived from
the ionization balance by 0.2--0.3 dex, and they are a better estimate for
dwarf and subgiants with $\log g > 3.5 $ when derived 
in the scale of Fuhrmann.

The applicability of the strong-line method to  metal-poor stars
becomes more difficult when the metallicity drops, as the wings of the lines
become less prominent. For the extreme metal-poor stars with 
[Fe/H] $< -3$, it is doubtful that any useful result can be drawn
 from this method (Fuhrmann 1998b).

\section{The connection between departures from the excitation equilibrium 
and the ionization balance.}

We mentioned in the introduction that Magain \& Zhao (1996) found strong
 departures from the excitation equilibrium for neutral iron lines in 
metal-poor stars. They presented a complicated picture in which
 the iron abundances derived from  high and low excitation  lines differ 
by up to +0.3 dex for the hotter stars in their sample ($T_{\rm eff} 
\sim$ 6000--6200 K). The estimates from  high  excitation lines were 
generally larger for almost all the studied stars, in particular for the stars 
in  the turnoff. 

If such a discrepancy is present, it will affect significantly  the
gravities derived through the ionization equilibrium. Assuming the
continuum as shaped by the H$^{-}$ bound-free opacity, for an element
which is mostly ionized (as it is the case for  iron in solar-like
stars),  we  expect the abundances derived from neutral  lines to have a
dependence with  temperature of $-1.3 \times 10^{-3}$ dex/K when the
excitation potential ($\chi$) is about 0 eV and $-0.4 \times 10^{-3}$
dex/K when  $\chi$ is about 5 eV
 (see Gray 1992, pages 286--287). Therefore,  a discrepancy of $+0.3$
 dex between high excitation  and low excitation lines correspond to a
systematic deviation in the derived  mean abundance of roughly $-0.3$
dex compared
 with the result that  would be obtained when the excitation balance is
 accomplished.  Approximating the dependence of the electron pressure
on the star's gravity as P$_{\rm e}  \propto g^{\frac{1}{3}}$,  the
ratio of the line to continuum opacities for a Fe~II line is expected
to be $\frac{l_{\nu}}{\chi_{\nu}} \propto g^{-\frac{1}{3}}$ (Gray 1992;
pages 287--290) and then, as the line opacity is also proportional to
the abundance N(Fe), the product  N(Fe) $ g^{-\frac{1}{3}}$ remains
constant for a given line equivalent width. In this situation, we can
estimate errors in the ionization balance gravities   induced by errors
in the abundance derived from Fe I lines  as {\nobreak $\Delta (\log g
$) = $ 3 \times \Delta $ [Fe/H]}. This has been illustrated in Fig. 9 for the
sake of clarity.  We are allowed then to translate the $-0.3$ dex error
in the iron abundance to a systematic deviation of $-0.9$ dex in the
iron ionization gravities, which is close to the discrepancies  in Fig.
7 for the more metal-poor stars in the samples of Fuhrmann et al.
(1997) and Magain (1989).

HD140283 is probably the most metal-poor star in this study, and it has
been analyzed by four different sources. Magain (1989) derived 
photometrically (neglecting the reddening) $T_{\rm eff} = 5640$ K for this
star, and then $\log g = 3.1$ dex from the iron ionization equilibrium. If
this star fits in the general panorama for departures from excitation
equilibrium described by Magain \& Zhao (1996; their Fig. 4), they 
would have found a discrepancy of $\sim +0.1$ dex between the iron 
abundances derived from high and low-excitation potential lines. The 
higher effective temperature ($+ 110$ K) employed by Ryan et al. (1996) will
help to achieve the Fe I excitation balance, and will result in a larger gravity
 (also metallicity), as is indeed the case: 
$\log g= 3.4$ dex. The  even slightly
hotter $T_{\rm eff}$   (between $+115$ and $+139$ K)  of Gratton et al. (1996)
might overcorrect  for the effect, as   confirmed by the 
small but {\it negative}  slope of Fe I abundances with excitation 
potential they found for a  selection of stars.  Nonetheless,
 we cannot explain the low spectroscopic (iron ionization) 
gravity found by Fuhrmann et al. (1997), $\log g = 3.2$ 
(in agreement with Magain 1989), as they assign to the star a much higher
 $T_{\rm eff}$ ($\sim +200 $K), but we note that their
 $f$-values have been estimated from the analysis of the solar flux
 spectrum with a flux-constant model atmosphere, in opposition to the
general rule followed  by Magain, Gratton et al.,  and Ryan et al., who
use laboratory data whenever possible.

We conclude that it is possible, if not likely, that the observed 
departures  from the excitation equilibrium in neutral iron lines are connected
with the discrepancies reported here between iron ionization gravities and
the trigonometric estimates. The  physical origin of both has to be 
investigated, and several paths to follow are suggested in the next section.

\section{Conclusions and suggestions for future work.}

The comparison of spectroscopic measurements and synthetic 
spectra can be used to constrain what the fundamental parameters 
characterizing a star should be. When one of these parameters can be
reliably measured through another means, it can  be used to carry 
out a consistency check,  testing the adequacy of the modeling.
Unfortunately, direct measurement of fundamental stellar parameters,
such as the radius, or the mass, is rarely possible, especially for
isolated stars. Additionally,  it is particularly difficult  to determine
 the distance to the star, which is needed to derive other 
basic properties, such as the luminosity. 

Measurements of stellar parallaxes provide distance
estimates for a set of nearby stars that has been significantly  enlarged
with the recent release of the {\it Hipparcos} mission data. Here, 
we have derived the distances for late-type stars with highly accurate
 {\it Hipparcos} parallaxes and which had been the subject of
high-resolution spectroscopic analysis in the literature. The
parallaxes with the V magnitude  provide the absolute 
magnitude M$_v$ which, combined with  models of stellar evolution and an
estimate of the effective temperature, gives a fairly accurate
determination of the gravity. The comparison of these {\it trigonometric}
gravities with those derived from published spectroscopic analyses
(ionization balance),  assuming the same effective temperatures, reveals:

\begin{itemize}
\item  good agreement for  stars  in the metallicity range 
$-1.0 < $ [Fe/H] $< 0.3$, the mean difference being $<\log g$ (Spec.) - 
$\log g$ (Trig.) $> = +0.10 \pm 0.17$ (1$\sigma$) dex, 
consistent with a scatter in the spectroscopic (iron ionization balance) 
gravities  of 0.15 dex.

\item a larger discrepancy for  stars with metallicities below [Fe/H] = --1. 
The discrepancy increases towards lower metallicities,  as  suggested by  
three independent sources for the spectroscopic gravities.
\end{itemize}

These findings cast a shadow upon analyses of extreme metal-poor 
stars based on gravities derived from the ionization balance, 
until independent estimates
of the stellar fundamental properties can be obtained. Future astrometric
missions, such as SIM (Unwin, Yu \& Shao 1997) or GAIA (Lindegren \& 
Perryman 1997) will push the accuracy of the parallaxes down to the level
of a few microarcseconds, and reach stars as faint as  V=15.

Our discussion cannot cover metal-poor low-gravity stars, as they
are not represented in the studied sample. However, we note that a similar 
consistency check that involves the absolute brightness of  RR Lyr 
variables as an  independent distance estimate,  
has been satisfactorily applied to  giants in the  globular clusters 
M15 (Sneden et al. 1997) and  NGC7006 (Kraft et al. 1998).

A comparison of the strong-line gravities derived by Edvardsson (1988)
and Fuhrmann (1988a) confirms  that this method provide systematically
larger gravities. However, while  the strong-line gravities get
closer to the physical ones for the stars in or near the turnoff, 
they are even further away than the iron ionization gravities 
for the low-gravity stars analyzed by Edvardsson (1988). The origin 
of these systematic differences  remains unclear. 
We  remark that this method cannot be applied to  extreme  
metal-poor stars,  as the wings of the lines lose their sensitivity to pressure.

Some of the  published claims on departures from the iron excitation 
equilibrium in metal-poor stars    appear to be connected with the 
discrepancies found here between spectroscopic (iron ionization) and
trigonometric gravities.

Several factors  influence the spectroscopically derived gravities and
are capable of producing systematic deviations, namely, the scale of
effective temperatures, the treatment of the line damping, the atomic
data,  and  other  ingredients involved in
 the modeling, as well as the measurement error itself.  Departures
from LTE are expected to make the spectroscopic gravities to diverge
further from the physical ones when the metallicity gets lower.  Very
recently, Th\'evenin \& Idiart (1999) have explored this possibility in
detail. Using complex atomic models for neutral and ionized iron and
photoionization cross-sections from the Iron Project (Bautista 1997),
they have studied departures from LTE in the line formation for a
sample of stars in the range $-4 \leq $ [Fe/H] $\leq 0$, concluding
that they exist and become significantly larger towards lower
metallicities as a result of  iron overionization compared to LTE
predictions. In Fig. 10, we overplotted their corrections to the LTE
gravities (filled stars; from their Table 1)  on top of our Fig. 4b, as
well as a  least-squares quadratic fit to their data (solid line).
Constrasting of Gratton et al's data (filled circles) with the
corrections of Th\'evenin \& Idiart  suggests that  NLTE in the line
formation is not the unique element at work, and something else is
distorting further the gravity estimates. Nevertheless, a very
different perspective emerges when the sample of Gratton et al.  is
excluded from the comparison. In that case, a quadratic polynomial fit
to the observed discrepancies (dashed line) will be roughly consistent
with the departures from LTE predicted by Th\'evenin \& Idiart.

Definitely, NLTE in the line formation may explain the discrepancies
between spectroscopic and trigonometric gravities  for metal-poor
dwarfs at about  [Fe/H] $\sim -2.5$.  Additional effects might present, 
as suggested by the fact that the slope of the discrepancies
between LTE iron ionization  and the trigonometric gravities for stars
more metal-poor than [Fe/H]=$-1$ is steeper ($\sim +0.35$; see \S 3.2)
 than Th\'evenin \& Idiart's predictions  ($+0.06$).  This conclusion
is independently reached using either the sample of Gratton et al.,
Fuhrmann et al., or Magain.  NLTE effects  could be important when
modeling the atmospheres of metal-poor stars.  Hauschildt, Allard, \&
Baron (1999) have computed flux-constant model atmospheres in NLTE for
Vega and the Sun. An extension of these computations to metal-poor
stars is desirable to test for departures from LTE  that  induce
deviations from  the ionization and excitation balance.  Besides, we
cannot exclude the effect of atmospheric inhomogeneities on the line
 formation as a factor inducing the observed discrepancies. Allende
 Prieto et al. (1995, 1999) compared convective line asymmetries in the
 metal-poor star HD140283 with the solar case, finding significantly
 larger velocities for the metal deficient  star. Detailed three
dimensional hydrodynamical simulations of surface convection in
metal-poor stars (Asplund et al. 1999) should provide tremendous
insight on such a controversial issue.

We still have to emphasize that systematic effects between different
analyses are  large. This reveals inadequate implementations of the
methods employed in the spectroscopic studies. These artificial
systematic deviations must be reduced to search for and clarify
possible failures either in  the physical assumptions or in the model
atmospheres.

\acknowledgements

We thank the anonymous
referee for many useful comments and suggestions.
This work has been partially funded by the Spanish DGES under projects
PB92-0434-C02-01 and PB95-1132-C02-01, the U.S. National Science
Foundation (grant AST961814),  and the Robert A. Welch Foundation of
Houston, Texas. We have made use of data from the {\it Hipparcos}
astrometric mission of the ESA,  and the NASA ADS.

\clearpage

\figcaption{Distribution of the nominal relative errors in the {\it Hipparcos} 
parallaxes for the program stars.}

\figcaption{The $T_{\rm eff}-M_v$ diagram. Selected
isochrones from Vandenberg (1985; solar metallicity) and 
Bergbusch \& Vandenberg (1992; metal-poor)
are drawn. Stars from Table 1 are
plotted. Filled circles represent stars of  [Fe/H] $\leq$ --0.47, while
open circles represent  stars with [Fe/H] $>$ --0.47.}

\figcaption{Comparison of spectroscopic and trigonometric
gravities. Metal-rich stars ([Fe/H] $> -0.47$) are indicated with 
filled circles and metal-poor stars are  with open circles. 
 The solid line  corresponds to the
case $\log g$\  (spectroscopic) = $\log g$\  (trigonometric).}

\figcaption{a) The differences between  spectroscopic and trigonometric
gravities are plotted as a function of the trigonometric gravity and
the effective temperature. The different symbols 
identify the different sources for the spectroscopic data: Gratton et al. (1996; 
filled circles), Edvardsson (1988; asterisks), Bonnell \& Bell (1993; pluses), 
Fuhrmann et al. (1997; rhombi), Magain (1989; squares), and 
Ryan et al. (1996; cross); b)  Differences between  spectroscopic and trigonometric gravities against  the iron abundance.}

\figcaption{Correlation between the relative errors in the parallaxes and
the stellar iron abundance (upper panel). Histogram of the metallicity 
distribution for the stars with relative errors 
$0.05 \le \sigma(\pi)/\pi \le 0.10$ (mid panel). Differences between 
spectroscopic and trigonometric gravities for the stars with relative errors
$0.05 \le \sigma(\pi)/\pi \le 0.10$ (lower panel). The symbols follow the 
same code  explained in Fig. 4.}

\figcaption{Differences between the  trigonometric gravities derived from:
Bertelli et al. (1994) or Bergbush \& Vandenberg (1992) (upper panel),
and Bergbush \& Vandenberg (1992) assuming all the stars 
have [Fe/H]=--0.47  or taking their real metallicities into account to estimate
their masses and bolometric corrections (lower panel). Note that 
the (OPAL) Bertelli et al's isochrones cover the range 
$-1.7 <$ [Fe/H] $< 0.4$, while Bergbush \& Vandenberg's (O-enhanced) 
models span the metallicity band $-2.3 <$ [Fe/H] $< -0.4$. The extreme
values were applied when the metallicity exceeded the limits.}

\figcaption{Differences between the spectroscopic (iron ionization) gravities
and the trigonometric estimates as a function of the stellar metallicity in the
 ranges  $-1.0 <$ [Fe/H] $< 0.3$ (a) and $-2.6 <$ [Fe/H] $< -1.0$ (b). The 
metallicities given by Magain (1989) have been shifted by $+0.17$ to
take into account that he adopted a higher solar abundance than the 
($\sim $ meteoritic) value used by the others. The solid lines show 
$\chi^2$-linear fits to the data for each reference.}

\figcaption{The strong-line gravities (open circles) and the iron ionization 
gravities (filled circles) are compared with the {\it Hipparcos} trigonometric
gravities. Symbols in the  upper right part of the plot correspond to the
spectroscopic analysis of Fuhrmann et al. (1997) and Fuhrmann (1998), while
those in the bottom left  correspond to the study of Edvardsson (1988).}

\figcaption{The figure illustrates the behavior of the abundances derived
from neutral iron lines (left panel)
 and singly ionized iron lines (right panel) 
in a solar-like photosphere. Besides, the arrows indicate the connection
between departures in the excitation balance and  those in the
 ionization equilibrium.}

\figcaption{Differences between  spectroscopic and trigonometric gravities against  the iron abundance. The different symbols 
identify the various sources for the spectroscopic data and are the same used
in previous figures (e.g. Fig. 4). The stars show  the NLTE corrections to
the ionization equilibrium gravities calculated by Th\'evenin \& Idiart 
(1999). The solid and dashed lines correspond to  least-square 
 polynomial fits to the corrections of Th\'evenin \& Idiart and the differences
$\log g$(Spec.) $- \log g$(Trig.) (excluding Gratton et al's data), respectively.}

\clearpage


\begin{deluxetable}{rrrrrrrrrrrrr}
\tiny
\tablecaption{Data for the stars in the comparison
\label{table1}}
\tablehead{
\colhead{Star}  & \colhead{[Fe/H]} & \colhead{Mass} & \colhead{V} & 
\colhead{M$_v$} &  \colhead{$T_{\rm eff}$} &
\colhead{$\log g$} & \colhead{$\log g$} & \colhead{$\pi$} &  \colhead{$\sigma$($\pi$)} & \colhead{Ref.\tablenotemark{a}} & \colhead{Iso.\tablenotemark{b}} & \colhead{CCDM\tablenotemark{c}} \nl
 &   &  \colhead{M$_\odot$} & & & \colhead{K} & \colhead{Hipparcos} & \colhead{Spectroscopic} & \colhead{mas}  & \colhead{mas} & & }
\startdata 
HR 17 & -0.34 & 1.01 & 6.21 & 3.61 & 6168 & 4.09 & 4.07 &   30.26 & 
    0.69 & 0 & 1 \nl
HR 33 & -0.37 & 1.02 & 4.89 & 3.51 & 6193 & 4.06 & 4.16 &   52.94 & 
    0.77 & 0 & 1 \nl
HR 35 & -0.14 & 1.02 & 5.24 & 3.55 & 6491 & 4.15 & 4.30 &   45.85 & 
    0.66 & 0 & 1 \nl
HR 107 & -0.37 & 1.04 & 6.05 & 3.25 & 6431 & 3.95 & 4.06 &   27.51 & 
    0.86 & 0 & 1 \nl
HR 140 & 0.04 & 1.02 & 5.57 & 3.53 & 6437 & 4.13 & 4.42 &   39.03 & 
    0.62 & 0 & 1 \nl
HR 145 & -0.23 & 1.02 & 6.32 & 3.44 & 6178 & 4.03 & 3.93 &   26.59 & 
    0.85 & 0 & 1 \nl
HR 203 & -0.25 & 1.01 & 6.15 & 3.63 & 5793 & 3.99 & 4.12 &   31.39 & 
    1.03 & 0 & 1 & 00455-1253 \nl
HR 219 & -0.31 & 0.92 & 3.46 & 4.59 & 5897 & 4.36 & 4.35 &  167.99 & 
    0.62 & 0 & 1 & 00491+5749 \nl
HR 235 & -0.20 & 0.96 & 5.17 & 4.22 & 6168 & 4.31 & 4.53 &   64.69 & 
    1.03 & 0 & 1 \nl
HR 244 & -0.04 & 1.02 & 4.80 & 3.46 & 6105 & 4.01 & 4.10 &   53.85 & 
    0.60 & 0 & 1 & 00531+6107 \nl
HR 340 & -0.13 & 1.05 & 5.67 & 2.06 & 5780 & 3.27 & 3.26 &   18.98 & 
    0.71 & 0 & 1 \nl
HR 366 & -0.33 & 1.04 & 5.14 & 3.20 & 6419 & 3.93 & 4.17 &   41.01 & 
    0.89 & 0 & 1 & 01144-0755 \nl
HR 370 & 0.09 & 0.98 & 4.97 & 4.08 & 5994 & 4.21 & 4.26 &   66.43 & 
    0.64 & 0 & 1 \nl
HR 448 & 0.13 & 1.03 & 5.75 & 3.39 & 5804 & 3.90 & 3.80 &   33.71 & 
    0.72 & 0 & 1 \nl
HR 458 & 0.06 & 1.02 & 4.10 & 3.45 & 6125 & 4.02 & 3.98 &   74.25 & 
    0.72 & 0 & 1 & 01367+4125 \nl
HR 483 & -0.04 & 0.94 & 4.96 & 4.45 & 5825 & 4.29 & 4.33 &   79.09 & 
    0.83 & 0 & 1 \nl
HR 573 & -0.36 & 0.98 & 6.10 & 4.00 & 6206 & 4.24 & 4.12 &   37.97 & 
    0.61 & 0 & 1 \nl
HR 646 & -0.30 & 1.04 & 5.23 & 2.84 & 6334 & 3.75 & 3.94 &   33.19 & 
    0.85 & 0 & 1 \nl
HR 672 & 0.02 & 1.01 & 5.60 & 3.61 & 5968 & 4.03 & 4.25 &   40.04 & 
    0.92 & 0 & 1 \nl
HR 720 & -0.18 & 1.02 & 5.89 & 3.45 & 5840 & 3.93 & 3.81 &   32.48 & 
    0.84 & 0 & 1 \nl
HR 740 & -0.24 & 1.04 & 4.74 & 2.68 & 6409 & 3.71 & 4.06 &   38.73 & 
    0.87 & 0 & 1 \nl
HR 784 & -0.01 & 0.97 & 5.79 & 4.12 & 6209 & 4.29 & 4.16 &   46.42 & 
    0.82 & 0 & 1 \nl
HR 799 & -0.04 & 1.00 & 4.10 & 3.85 & 6239 & 4.20 & 4.35 &   89.03 & 
    0.79 & 0 & 1 & 02441+4913 \nl
HR 962 & 0.07 & 1.04 & 5.07 & 3.32 & 6015 & 3.87 & 3.94 &   44.69 & 
    0.75 & 0 & 1 & 03128-0112 \nl
HR 1083 & -0.14 & 1.04 & 4.71 & 3.05 & 6695 & 3.94 & 4.33 &   46.65 & 
    0.48 & 0 & 1 & 03294-6256 \nl
HR 1101 & -0.11 & 1.01 & 4.29 & 3.60 & 5950 & 4.02 & 4.00 &   72.89 & 
    0.78 & 0 & 1 \nl
HR 1173 & 0.01 & 1.04 & 4.22 & 2.95 & 6569 & 3.87 & 4.20 &   55.79 & 
    0.69 & 0 & 1 \nl
HR 1257 & 0.03 & 1.04 & 5.36 & 2.66 & 6219 & 3.65 & 3.85 &   28.87 & 
    0.82 & 0 & 1 \nl
HR 1294 & -0.15 & 0.93 & 6.37 & 4.54 & 5727 & 4.29 & 4.28 &   43.12 & 
    0.50 & 0 & 1 \nl
HR 1489 & 0.03 & 1.02 & 5.97 & 3.50 & 5942 & 3.98 & 4.01 &   32.03 & 
    0.91 & 0 & 1 \nl
HR 1536 & 0.14 & 1.01 & 5.77 & 3.65 & 5807 & 4.00 & 3.98 &   37.73 & 
    0.89 & 0 & 1 \nl
HR 1545 & -0.37 & 1.02 & 6.27 & 3.53 & 6342 & 4.11 & 4.55 &   28.28 & 
    0.80 & 0 & 1 \nl
HR 1673 & -0.32 & 1.04 & 5.11 & 3.12 & 6397 & 3.89 & 4.00 &   39.99 & 
    0.70 & 0 & 1 \nl
HR 1687 & 0.20 & 1.04 & 5.89 & 2.97 & 6456 & 3.84 & 4.02 &   26.04 & 
    0.83 & 0 & 1 \nl
HR 1729 & -0.04 & 0.97 & 4.69 & 4.18 & 5824 & 4.20 & 4.26 &   79.08 & 
    0.90 & 0 & 1 & 05192+4007 \nl
HR 1780 & 0.01 & 0.97 & 5.00 & 4.17 & 6108 & 4.27 & 4.07 &   68.19 & 
    0.94 & 0 & 1 & 05244+1723 \nl
HR 1907 & -0.50 & 0.94 & 4.09 & 1.33 & 4720 & 2.55 & 2.30 &   28.10 & 
    0.91 & 2 & 0 \nl
HR 1983 & -0.12 & 1.00 & 3.59 & 3.83 & 6299 & 4.21 & 4.10 &  111.49 & 
    0.60 & 0 & 1 & 05445-2226 \nl
HR 2047 & -0.04 & 0.91 & 4.39 & 4.70 & 5895 & 4.40 & 4.21 &  115.43 & 
    1.08 & 0 & 1 & 05544+2017 \nl
HR 2141 & -0.24 & 0.98 & 6.12 & 4.01 & 5887 & 4.15 & 4.35 &   37.90 & 
    0.83 & 0 & 1 & 06061+3524 \nl
HR 2220 & -0.01 & 1.01 & 5.20 & 3.58 & 6467 & 4.16 & 4.30 &   47.33 & 
    0.86 & 0 & 1 & 06148+1909 \nl
HR 2354 & 0.14 & 0.99 & 6.45 & 3.94 & 5739 & 4.08 & 4.10 &   31.46 & 
    0.52 & 0 & 1 \nl
HR 2493 & -0.37 & 0.95 & 6.43 & 4.31 & 6037 & 4.30 & 4.59 &   37.60 & 
    0.65 & 0 & 1 \nl
HR 2548 & -0.24 & 1.04 & 5.14 & 3.13 & 6386 & 3.89 & 4.14 &   39.66 & 
    0.53 & 0 & 1 \nl
HR 2601 & -0.50 & 0.94 & 6.20 & 2.82 & 6056 & 3.62 & 4.02 &   21.10 & 
    0.90 & 0 & 0 & 06588+2605 \nl
HR 2721 & -0.27 & 0.94 & 5.54 & 4.41 & 5913 & 4.30 & 4.38 &   59.31 & 
    0.69 & 0 & 1 \nl
HR 2835 & -0.50 & 0.91 & 6.54 & 4.09 & 6227 & 4.25 & 4.30 &   32.43 & 
    0.91 & 0 & 0 \nl
HR 2883 & -0.70 & 0.91 & 5.90 & 3.52 & 5994 & 3.90 & 4.15 &   33.40 & 
    0.93 & 0 & 0 & 07321-0853 \nl
HR 2906 & -0.18 & 1.05 & 4.44 & 2.39 & 6149 & 3.52 & 3.75 &   38.91 & 
    0.66 & 0 & 1 \nl
HR 2943 & -0.06 & 1.04 & 0.40 & 2.68 & 6605 & 3.76 & 4.13 &  285.93 & 
    0.88 & 0 & 1 & 07393+0514 \nl
HR 3018 & -0.72 & 0.86 & 5.36 & 4.45 & 5890 & 4.27 & 4.42 &   65.79 & 
    0.56 & 0 & 0 \nl
HR 3018 & -0.78 & 0.85 & 5.36 & 4.45 & 5723 & 4.22 & 4.21 &   65.79 & 
    0.56 & 0 & 0 \nl
HR 3176 & 0.09 & 1.02 & 5.30 & 3.46 & 5728 & 3.90 & 3.84 &   42.86 & 
    0.97 & 0 & 1 \nl
HR 3220 & -0.29 & 1.04 & 4.74 & 3.09 & 6457 & 3.89 & 4.28 &   46.75 & 
    3.38 & 0 & 1 \nl
HR 3262 & -0.26 & 1.00 & 5.13 & 3.84 & 6301 & 4.21 & 4.11 &   55.17 & 
    0.93 & 0 & 1 \nl
HR 3538 & 0.02 & 0.90 & 6.01 & 4.85 & 5687 & 4.38 & 4.37 &   58.50 & 
    0.88 & 0 & 1 \nl
HR 3578 & -0.76 & 0.87 & 5.80 & 4.16 & 5971 & 4.18 & 4.62 &   46.90 & 
    0.97 & 0 & 0 \nl
HR 3648 & -0.06 & 1.00 & 5.18 & 3.72 & 5830 & 4.03 & 3.93 &   51.12 & 
    0.72 & 0 & 1 & 09143+6125 \nl
HR 3775 & -0.21 & 1.05 & 3.17 & 2.52 & 6296 & 3.61 & 3.80 &   74.15 & 
    0.74 & 0 & 1 & 09329+5141 \nl
HR 3881 & 0.07 & 1.00 & 5.08 & 3.75 & 5828 & 4.04 & 3.92 &   54.26 & 
    0.74 & 0 & 1 \nl
HR 3951 & 0.11 & 0.93 & 5.37 & 4.50 & 5675 & 4.26 & 4.28 &   67.14 & 
    0.83 & 0 & 1 & 10010+3155 \nl
HR 3954 & 0.01 & 1.05 & 5.71 & 2.34 & 6292 & 3.53 & 3.84 &   21.15 & 
    0.72 & 0 & 1 \nl
HR 4012 & 0.15 & 1.05 & 6.02 & 2.44 & 6023 & 3.50 & 3.64 &   19.27 & 
    0.89 & 0 & 1 \nl
HR 4027 & -0.01 & 0.99 & 6.46 & 3.93 & 5772 & 4.09 & 4.11 &   31.25 & 
    0.81 & 0 & 1 \nl
HR 4039 & -0.38 & 0.98 & 5.81 & 4.03 & 6143 & 4.23 & 4.54 &   44.01 & 
    0.75 & 0 & 1 & 10172+2306 \nl
HR 4067 & 0.14 & 1.04 & 5.73 & 2.78 & 6186 & 3.69 & 3.83 &   25.65 & 
    0.70 & 0 & 1 \nl
HR 4150 & -0.21 & 1.04 & 6.29 & 2.96 & 6465 & 3.84 & 4.05 &   21.60 & 
    0.75 & 0 & 1 \nl
HR 4158 & -0.24 & 1.00 & 5.71 & 3.76 & 6093 & 4.12 & 4.18 &   40.67 & 
    0.68 & 0 & 1 & 10365-1214 \nl
HR 4257 & -0.01 & 1.05 & 3.78 & 1.42 & 5025 & 2.75 & 2.95 &   33.71 & 
    0.57 & 1 & 1 & 10535-5851 \nl
HR 4257 & 0.00 & 1.05 & 3.78 & 1.42 & 4870 & 2.70 & 3.01 &   33.71 & 
    0.57 & 2 & 1 & 10535-5851 \nl
HR 4277 & 0.00 & 0.96 & 5.03 & 4.29 & 5811 & 4.23 & 4.09 &   71.04 & 
    0.66 & 0 & 1 \nl
HR 4285 & -0.25 & 1.04 & 6.03 & 2.74 & 5853 & 3.58 & 3.74 &   21.98 & 
    0.75 & 0 & 1 & 11003+4255 \nl
HR 4287 & 0.14 & 1.05 & 4.08 & 0.44 & 4820 & 2.23 & 2.60 &   18.71 & 
    1.03 & 1 & 1 \nl
HR 4395 & -0.15 & 1.05 & 5.08 & 1.87 & 6525 & 3.40 & 3.63 &   22.80 & 
    0.84 & 0 & 1 \nl
HR 4421 & -0.51 & 0.93 & 5.83 & 3.24 & 6623 & 3.95 & 4.23 &   30.40 & 
    0.60 & 0 & 0 \nl
HR 4529 & 0.16 & 1.04 & 6.24 & 3.09 & 5977 & 3.76 & 3.80 &   23.49 & 
    0.80 & 0 & 1 & 11484-1018 \nl
HR 4540 & 0.10 & 1.02 & 3.59 & 3.40 & 6065 & 3.98 & 4.06 &   91.74 & 
    0.77 & 0 & 1 & 11507+0146 \nl
HR 4657 & -0.66 & 0.87 & 6.11 & 4.34 & 6267 & 4.34 & 4.60 &   44.34 & 
    1.01 & 0 & 0 & 12152-1019 \nl
HR 4688 & 0.19 & 1.04 & 6.38 & 2.90 & 6256 & 3.76 & 4.01 &   20.12 & 
    0.77 & 0 & 1 \nl
HR 4734 & 0.12 & 0.97 & 6.25 & 4.12 & 5665 & 4.13 & 4.08 &   37.50 & 
    0.72 & 0 & 1 \nl
HR 4767 & -0.13 & 0.93 & 6.20 & 4.49 & 5998 & 4.35 & 4.37 &   45.58 & 
    0.62 & 0 & 1 \nl
HR 4785 & -0.19 & 0.92 & 4.24 & 4.63 & 5814 & 4.35 & 4.29 &  119.46 & 
    0.83 & 0 & 1 \nl
HR 4845 & -0.51 & 0.86 & 5.95 & 4.75 & 5868 & 4.38 & 4.15 &   57.57 & 
    0.64 & 0 & 0 \nl
HR 4903 & 0.29 & 1.04 & 5.89 & 2.89 & 5880 & 3.65 & 3.78 &   25.17 & 
    0.76 & 0 & 1 \nl
HR 4981 & -0.20 & 1.05 & 5.04 & 2.48 & 6301 & 3.60 & 3.78 &   30.72 & 
    0.80 & 0 & 1 & 13120-1611 \nl
HR 4983 & 0.00 & 0.94 & 4.23 & 4.42 & 5952 & 4.32 & 4.31 &  109.23 & 
    0.72 & 0 & 1 & 13118+2753 \nl
HR 4989 & -0.29 & 1.01 & 4.90 & 3.62 & 6263 & 4.12 & 4.32 &   55.49 & 
    0.65 & 0 & 1 & 13142-5906 \nl
HR 5011 & 0.10 & 0.99 & 5.19 & 3.92 & 5920 & 4.13 & 3.96 &   55.71 & 
    0.85 & 0 & 1 & 13168+0925 \nl
HR 5019 & -0.03 & 0.87 & 4.74 & 5.09 & 5552 & 4.42 & 4.33 &  117.30 & 
    0.71 & 0 & 1 & 13185-1818 \nl
HR 5235 & 0.20 & 1.05 & 2.68 & 2.41 & 5943 & 3.47 & 3.38 &   88.17 & 
    0.75 & 0 & 1 & 13547+1824 \nl
HR 5287 & 0.12 & 1.05 & 3.25 & 0.79 & 4620 & 2.32 & 2.20 &   32.17 & 
    0.77 & 1 & 1 \nl
HR 5323 & 0.03 & 1.04 & 5.53 & 2.92 & 6130 & 3.73 & 3.99 &   30.06 & 
    0.74 & 0 & 1 \nl
HR 5338 & -0.11 & 1.05 & 4.07 & 2.42 & 6127 & 3.52 & 3.71 &   46.74 & 
    0.87 & 0 & 1 \nl
HR 5353 & 0.19 & 0.99 & 6.47 & 3.89 & 5483 & 3.99 & 3.90 &   30.47 & 
    1.00 & 0 & 1 & 14180-0732 \nl
HR 5423 & 0.15 & 0.93 & 6.36 & 4.50 & 5545 & 4.22 & 3.88 &   42.43 & 
    0.59 & 0 & 1 \nl
HR 5447 & -0.41 & 1.02 & 4.47 & 3.52 & 6734 & 4.21 & 4.31 &   64.66 & 
    0.72 & 0 & 1 & 14347+2945 \nl
HR 5542 & 0.09 & 1.04 & 6.30 & 3.29 & 5896 & 3.82 & 3.98 &   24.99 & 
    0.80 & 0 & 1 \nl
HR 5691 & -0.02 & 1.04 & 5.15 & 3.13 & 6077 & 3.81 & 3.94 &   39.51 & 
    0.47 & 0 & 1 \nl
HR 5698 & -0.01 & 1.05 & 4.99 & 2.32 & 6245 & 3.52 & 3.76 &   29.27 & 
    0.76 & 0 & 1 \nl
HR 5723 & -0.16 & 1.05 & 4.92 & 2.37 & 6416 & 3.58 & 3.80 &   30.90 & 
    0.99 & 0 & 1 \nl
HR 5868 & -0.04 & 0.98 & 4.42 & 4.07 & 5847 & 4.16 & 4.06 &   85.08 & 
    0.80 & 0 & 1 \nl
HR 5908 & -0.06 & 1.05 & 4.13 & 0.64 & 4865 & 2.34 & 2.60 &   20.02 & 
    0.88 & 1 & 1 \nl
HR 5908 & 0.00 & 1.05 & 4.13 & 0.64 & 4780 & 2.31 & 2.49 &   20.02 & 
    0.88 & 2 & 1 \nl
HR 5914 & -0.46 & 1.01 & 4.60 & 3.60 & 5831 & 3.99 & 3.96 &   63.08 & 
    0.54 & 0 & 1 \nl
HR 5933 & -0.18 & 1.01 & 3.85 & 3.62 & 6268 & 4.12 & 4.04 &   89.92 & 
    0.72 & 0 & 1 & 15564+1540 \nl
HR 5968 & -0.22 & 0.97 & 5.39 & 4.18 & 5745 & 4.17 & 4.11 &   57.38 & 
    0.71 & 0 & 1 & 16011+3318 \nl
HR 5996 & 0.27 & 0.98 & 6.32 & 4.02 & 5756 & 4.12 & 4.10 &   34.60 & 
    1.00 & 0 & 1 \nl
HR 6189 & -0.53 & 0.93 & 6.33 & 3.14 & 6170 & 3.78 & 4.12 &   23.02 & 
    0.89 & 0 & 0 \nl
HR 6202 & -0.32 & 1.05 & 5.55 & 2.41 & 6465 & 3.61 & 4.02 &   23.50 & 
    1.05 & 0 & 1 & 16419-1955 \nl
HR 6243 & -0.03 & 1.05 & 4.64 & 1.80 & 6361 & 3.33 & 3.55 &   27.04 & 
    1.08 & 0 & 1 \nl
HR 6315 & -0.18 & 0.98 & 4.88 & 3.99 & 6215 & 4.24 & 4.01 &   66.28 & 
    0.48 & 0 & 1 \nl
HR 6409 & 0.02 & 1.05 & 5.53 & 2.01 & 6233 & 3.38 & 3.54 &   19.80 & 
    0.72 & 0 & 1 \nl
HR 6458 & -0.38 & 0.92 & 5.38 & 4.59 & 5633 & 4.28 & 4.12 &   69.48 & 
    0.56 & 0 & 1 & 17206+3229 \nl
HR 6541 & -0.21 & 1.04 & 5.65 & 2.83 & 6196 & 3.71 & 3.69 &   27.26 & 
    0.73 & 0 & 1 \nl
HR 6569 & -0.31 & 1.04 & 4.76 & 3.06 & 6601 & 3.92 & 3.99 &   45.72 & 
    0.79 & 0 & 1 \nl
HR 6598 & -0.33 & 1.04 & 6.33 & 3.19 & 5757 & 3.74 & 3.68 &   23.53 & 
    0.54 & 0 & 1 \nl
HR 6649 & -0.35 & 1.01 & 6.19 & 3.62 & 6060 & 4.06 & 4.19 &   30.55 & 
    0.90 & 0 & 1 \nl
HR 6701 & -0.20 & 1.05 & 5.02 & 2.49 & 6129 & 3.55 & 3.75 &   31.13 & 
    0.47 & 0 & 1 \nl
HR 6775 & -0.54 & 0.90 & 5.05 & 4.08 & 6001 & 4.17 & 4.21 &   63.88 & 
    0.55 & 0 & 0 & 18071+3034 \nl
HR 6850 & -0.32 & 1.04 & 4.99 & 3.14 & 6525 & 3.93 & 4.05 &   42.56 & 
    0.45 & 0 & 1 \nl
HR 6869 & -0.10 & 1.05 & 3.23 & 1.84 & 4930 & 2.90 & 3.12 &   52.81 & 
    0.75 & 1 & 1 & 18214-0253 \nl
HR 6869 & 0.00 & 1.05 & 3.23 & 1.84 & 4950 & 2.91 & 2.94 &   52.81 & 
    0.75 & 2 & 1 & 18214-0253 \nl
HR 6907 & 0.07 & 1.04 & 5.90 & 3.10 & 6245 & 3.84 & 4.08 &   27.53 & 
    0.91 & 0 & 1 \nl
HR 6913 & 0.09 & 1.05 & 2.82 & 0.95 & 4775 & 2.45 & 2.73 &   42.20 & 
    0.90 & 1 & 1 \nl
HR 7061 & -0.09 & 1.04 & 4.19 & 2.79 & 6301 & 3.72 & 3.93 &   52.37 & 
    0.68 & 0 & 1 & 18457+2033 \nl
HR 7126 & 0.15 & 1.04 & 5.56 & 3.03 & 6517 & 3.89 & 4.22 &   31.24 & 
    0.75 & 0 & 1 \nl
HR 7232 & 0.05 & 0.88 & 6.15 & 4.98 & 5578 & 4.39 & 4.23 &   58.24 & 
    0.91 & 0 & 1 \nl
HR 7322 & -0.29 & 1.04 & 6.02 & 2.88 & 6254 & 3.75 & 3.90 &   23.57 & 
    0.57 & 0 & 1 \nl
HR 7534 & -0.16 & 1.02 & 5.00 & 3.40 & 6289 & 4.04 & 4.10 &   47.94 & 
    0.54 & 0 & 1 & 19465+3343 \nl
HR 7560 & 0.03 & 1.01 & 5.12 & 3.68 & 6047 & 4.08 & 4.22 &   51.57 & 
    0.77 & 0 & 1 & 19510+1025 \nl
HR 7602 & -0.13 & 1.04 & 3.71 & 3.03 & 5225 & 3.50 & 3.65 &   72.95 & 
    0.83 & 1 & 1 & 19553+0625 \nl
HR 7766 & -0.36 & 1.01 & 6.26 & 3.71 & 5948 & 4.06 & 4.07 &   30.84 & 
    0.88 & 0 & 1 \nl
HR 7875 & -0.40 & 1.04 & 5.11 & 3.19 & 6031 & 3.82 & 4.05 &   41.33 & 
    0.73 & 0 & 1 \nl
HR 7955 & 0.07 & 1.05 & 4.52 & 2.35 & 6064 & 3.48 & 3.81 &   36.87 & 
    0.46 & 0 & 1 & 20454+5735 \nl
HR 8027 & -0.37 & 1.02 & 5.76 & 3.48 & 6246 & 4.06 & 4.27 &   35.07 & 
    0.83 & 0 & 1 \nl
HR 8041 & 0.13 & 0.97 & 6.21 & 4.10 & 5734 & 4.14 & 3.83 &   37.80 & 
    1.01 & 0 & 1 \nl
HR 8077 & -0.08 & 1.04 & 5.94 & 3.10 & 6095 & 3.80 & 3.90 &   27.06 & 
    1.07 & 0 & 1 & 21054+0558 \nl
HR 8181 & -0.62 & 0.87 & 4.21 & 4.39 & 6244 & 4.35 & 4.62 &  108.50 & 
    0.59 & 0 & 0 \nl
HR 8354 & -0.59 & 0.92 & 5.52 & 3.31 & 6378 & 3.91 & 4.11 &   36.15 & 
    0.69 & 0 & 0 \nl
HR 8472 & 0.03 & 1.05 & 5.24 & 2.40 & 6160 & 3.52 & 3.76 &   27.03 & 
    0.49 & 0 & 1 & 22119+5650 \nl
HR 8665 & -0.30 & 1.04 & 4.20 & 3.15 & 6184 & 3.84 & 3.92 &   61.54 & 
    0.77 & 0 & 1 & 22467+1211 \nl
HR 8697 & -0.23 & 1.04 & 5.16 & 3.02 & 6250 & 3.81 & 4.14 &   37.25 & 
    0.76 & 0 & 1 \nl
HR 8729 & 0.08 & 0.93 & 5.45 & 4.52 & 5669 & 4.27 & 4.06 &   65.10 & 
    0.76 & 0 & 1 \nl
HR 8805 & -0.17 & 1.04 & 5.68 & 3.02 & 6502 & 3.87 & 4.08 &   29.33 & 
    0.62 & 0 & 1 \nl
HR 8853 & -0.02 & 0.98 & 5.58 & 4.04 & 6067 & 4.22 & 4.08 &   49.31 & 
    0.58 & 0 & 1 & 23167+5313 \nl
HR 8885 & -0.01 & 1.04 & 5.77 & 2.63 & 6443 & 3.70 & 3.78 &   23.59 & 
    0.77 & 0 & 1 & 23208+3810 \nl
HR 8969 & -0.17 & 1.02 & 4.13 & 3.43 & 6198 & 4.03 & 4.28 &   72.51 & 
    0.88 & 0 & 1 & 23399+0538 \nl
HD 400 & -0.26 & 1.01 & 6.21 & 3.61 & 6156 & 4.09 & 3.98 &   30.26 & 
    0.69 & 3 & 1 \nl
HD 2615 & -0.54 & 0.93 & 7.62 & 3.22 & 6324 & 3.86 & 4.11 &   13.20 & 
    0.93 & 0 & 0 \nl
HD 6434 & -0.53 & 0.86 & 7.72 & 4.69 & 5796 & 4.34 & 4.42 &   24.80 & 
    0.89 & 0 & 0 \nl
HD 14938 & -0.33 & 1.02 & 7.20 & 3.55 & 6180 & 4.07 & 4.04 &   18.58 & 
    1.11 & 0 & 1 \nl
HD 17548 & -0.54 & 0.87 & 8.16 & 4.54 & 6015 & 4.35 & 4.46 &   18.90 & 
    1.38 & 0 & 0 \nl
HD 18768 & -0.51 & 0.93 & 6.72 & 3.40 & 5769 & 3.77 & 4.00 &   21.65 & 
    0.82 & 0 & 0 \nl
HD 19445 & -1.97 & 0.75 & 8.04 & 5.10 & 6040 & 4.51 & 4.12 &   25.85 & 
    1.14 & 3 & 0 \nl
HD 19445 & -2.30 & 0.75 & 8.04 & 5.10 & 5880 & 4.46 & 4.10 &   25.85 & 
    1.14 & 4 & 1 \nl
HD 19445 & -1.88 & 0.76 & 8.04 & 5.10 & 6080 & 4.52 & 4.72 &   25.85 & 
    1.14 & 0 & 0 \nl
HD 19445 & -1.89 & 0.76 & 8.04 & 5.10 & 6052 & 4.51 & 4.78 &   25.85 & 
    1.14 & 0 & 0 \nl
HD 22879 & -0.76 & 0.83 & 6.68 & 4.75 & 5926 & 4.38 & 4.57 &   41.07 & 
    0.86 & 0 & 0 \nl
HD 25329 & -1.69 & 0.58 & 8.51 & 7.18 & 4849 & 4.77 & 4.65 &   54.14 & 
    1.08 & 0 & 0 \nl
HD 25704 & -0.79 & 0.85 & 8.11 & 4.51 & 5903 & 4.29 & 4.49 &   19.02 & 
    0.87 & 0 & 0 & 04017-5712 \nl
HD 30649 & -0.46 & 0.93 & 6.94 & 4.56 & 5789 & 4.32 & 4.32 &   33.44 & 
    1.12 & 0 & 1 & 04518+4550 \nl
HD 30743 & -0.42 & 1.02 & 6.27 & 3.53 & 6295 & 4.09 & 3.73 &   28.28 & 
    0.80 & 3 & 1 \nl
HD 30743 & -0.48 & 0.93 & 6.27 & 3.53 & 6288 & 3.98 & 3.80 &   28.28 & 
    0.80 & 3 & 0 \nl
HD 34328 & -1.90 & 0.74 & 9.43 & 5.24 & 5830 & 4.50 & 4.25 &   14.55 & 
    1.01 & 4 & 0 \nl
HD 34328 & -1.44 & 0.76 & 9.43 & 5.24 & 5986 & 4.55 & 4.89 &   14.55 & 
    1.01 & 0 & 0 \nl
HD 38007 & -0.30 & 1.02 & 6.85 & 3.55 & 5671 & 3.92 & 4.04 &   21.89 & 
    0.88 & 0 & 1 \nl
HD 43947 & -0.27 & 0.94 & 6.61 & 4.41 & 5966 & 4.32 & 4.51 &   36.32 & 
    0.90 & 0 & 1 \nl
HD 51929 & -0.59 & 0.87 & 7.39 & 4.51 & 5876 & 4.29 & 4.47 &   26.58 & 
    0.55 & 0 & 0 \nl
HD 59984 & -0.64 & 0.92 & 5.90 & 3.52 & 5911 & 3.87 & 4.06 &   33.40 & 
    0.93 & 0 & 0 & 07321-0853 \nl
HD 61421 & 0.01 & 1.04 & 0.40 & 2.68 & 6470 & 3.73 & 3.58 &  285.93 & 
    0.88 & 3 & 1 & 07393+0514 \nl
HD 62301 & -0.60 & 0.90 & 6.74 & 4.07 & 5948 & 4.15 & 4.19 &   29.22 & 
    0.96 & 0 & 0 \nl
HD 62644 & 0.03 & 1.04 & 5.04 & 3.13 & 5294 & 3.56 & 3.41 &   41.43 & 
    0.81 & 0 & 1 \nl
HD 64090 & -1.60 & 0.68 & 8.27 & 6.01 & 5515 & 4.66 & 4.99 &   35.29 & 
    1.04 & 0 & 0 & 07536+3037 \nl
HD 64606 & -0.93 & 0.72 & 7.43 & 6.01 & 5206 & 4.57 & 4.57 &   52.01 & 
    1.85 & 0 & 0 & 07546-0125 \nl
HD 66573 & -0.53 & 0.84 & 7.26 & 4.91 & 5739 & 4.39 & 4.42 &   33.88 & 
    1.17 & 0 & 0 \nl
HD 68284 & -0.55 & 0.93 & 7.75 & 3.40 & 5912 & 3.81 & 4.02 &   13.47 & 
    1.17 & 0 & 0 \nl
HD 69611 & -0.55 & 0.89 & 7.74 & 4.29 & 5808 & 4.20 & 4.47 &   20.46 & 
    1.16 & 0 & 0 \nl
HD 74011 & -0.57 & 0.90 & 7.42 & 4.08 & 5776 & 4.11 & 4.15 &   21.51 & 
    0.95 & 0 & 0 \nl
HD 78558 & -0.40 & 0.94 & 7.29 & 4.47 & 5736 & 4.27 & 4.41 &   27.27 & 
    0.91 & 0 & 1 \nl
HD 78747 & -0.62 & 0.85 & 7.72 & 4.72 & 5830 & 4.35 & 4.45 &   25.16 & 
    0.68 & 0 & 0 \nl
HD 84937 & -2.04 & 0.83 & 8.33 & 3.80 & 6344 & 4.13 & 4.06 &   12.44 & 
    1.06 & 0 & 0 \nl
HD 84937 & -2.10 & 0.83 & 8.33 & 3.80 & 6357 & 4.13 & 4.14 &   12.44 & 
    1.06 & 0 & 0 \nl
HD 84937 & -2.40 & 0.82 & 8.33 & 3.80 & 6200 & 4.09 & 3.60 &   12.44 & 
    1.06 & 4 & 1 \nl
HD 91324 & -0.23 & 1.04 & 4.89 & 3.19 & 6150 & 3.85 & 3.89 &   45.72 & 
    0.51 & 0 & 1 & 10314-5343 \nl
HD 91347 & -0.45 & 0.91 & 7.50 & 4.72 & 5901 & 4.40 & 4.28 &   27.79 & 
    0.82 & 0 & 1 \nl
HD 94028 & -1.38 & 0.80 & 8.21 & 4.63 & 6060 & 4.36 & 4.54 &   19.23 & 
    1.13 & 0 & 0 \nl
HD 98553 & -0.41 & 0.89 & 7.54 & 4.89 & 5905 & 4.46 & 4.58 &   29.47 & 
    0.89 & 0 & 1 \nl
HD 102365 & -0.08 & 0.87 & 4.89 & 5.06 & 5637 & 4.44 & 4.45 &  108.23 & 
    0.70 & 0 & 1 \nl
HD 108177 & -1.55 & 0.78 & 9.66 & 4.86 & 6178 & 4.47 & 4.50 &   10.95 & 
    1.29 & 0 & 0 \nl
HD 114762 & -0.67 & 0.88 & 7.30 & 4.26 & 5941 & 4.22 & 4.17 &   24.65 & 
    1.44 & 0 & 0 \nl
HD 114946 & 0.12 & 1.05 & 5.31 & 2.38 & 5198 & 3.22 & 3.72 &   25.89 & 
    0.73 & 0 & 1 \nl
HD 116064 & -1.86 & 0.78 & 8.80 & 4.76 & 5983 & 4.37 & 4.58 &   15.54 & 
    1.44 & 0 & 0  & 13217-3919 \nl
HD 116064 & -2.30 & 0.78 & 8.80 & 4.76 & 5850 & 4.33 & 3.70 &   15.54 & 
    1.44 & 4 & 1  & 13217-3919 \nl
HD 121384 & -0.34 & 1.04 & 6.00 & 3.09 & 5199 & 3.52 & 3.70 &   26.24 & 
    0.67 & 0 & 1 & 13565-5442 \nl
HD 126512 & -0.56 & 0.91 & 7.27 & 3.91 & 5784 & 4.05 & 3.97 &   21.32 & 
    0.83 & 0 & 0 \nl
HD 126681 & -1.09 & 0.73 & 9.28 & 5.69 & 5625 & 4.60 & 4.95 &   19.16 & 
    1.44 & 0 & 0 \nl
HD 130551 & -0.55 & 0.92 & 7.16 & 3.76 & 6307 & 4.14 & 4.40 &   20.94 & 
    0.88 & 0 & 0 \nl
HD 134169 & -0.68 & 0.90 & 7.67 & 3.80 & 5872 & 4.02 & 4.18 &   16.80 & 
    1.11 & 0 & 0 \nl
HD 134169 & -0.76 & 0.89 & 7.67 & 3.80 & 5887 & 4.02 & 4.30 &   16.80 & 
    1.11 & 0 & 0 \nl
HD 136352 & -0.21 & 0.90 & 5.65 & 4.83 & 5695 & 4.38 & 4.54 &   68.70 & 
    0.79 & 0 & 1 \nl
HD 140283 & -2.38 & 0.84 & 7.20 & 3.41 & 5755 & 3.81 & 3.66 &   17.44 & 
    0.97 & 0 & 1 \nl
HD 140283 & -2.41 & 0.84 & 7.20 & 3.41 & 5779 & 3.81 & 3.79 &   17.44 & 
    0.97 & 0 & 1 \nl
HD 140283 & -2.42 & 0.84 & 7.20 & 3.41 & 5763 & 3.81 & 3.60 &   17.44 & 
    0.97 & 0 & 1 \nl
HD 140283 & -2.34 & 0.84 & 7.20 & 3.41 & 5843 & 3.83 & 3.20 &   17.44 & 
    0.97 & 3 & 1 \nl
HD 140283 & -2.54 & 0.84 & 7.20 & 3.41 & 5750 & 3.80 & 3.40 &   17.44 & 
    0.97 & 5 & 1 \nl
HD 140283 & -2.70 & 0.84 & 7.20 & 3.41 & 5640 & 3.77 & 3.10 &   17.44 & 
    0.97 & 4 & 1 \nl
HD 144172 & -0.44 & 1.04 & 6.81 & 3.23 & 6324 & 3.92 & 4.10 &   19.25 & 
    0.90 & 0 & 1 \nl
HD 148211 & -0.61 & 0.89 & 7.69 & 4.08 & 5921 & 4.15 & 4.35 &   18.98 & 
    0.98 & 0 & 0 \nl
HD 148816 & -0.68 & 0.88 & 7.27 & 4.20 & 5928 & 4.19 & 4.39 &   24.34 & 
    0.90 & 0 & 0 \nl
HD 155358 & -0.61 & 0.89 & 7.28 & 4.09 & 5914 & 4.15 & 4.09 &   23.04 & 
    0.69 & 0 & 0 \nl
HD 157089 & -0.51 & 0.91 & 6.95 & 4.01 & 5833 & 4.10 & 4.35 &   25.88 & 
    0.95 & 0 & 0 \nl
HD 159307 & -0.68 & 0.92 & 7.40 & 3.04 & 6278 & 3.77 & 4.03 &   13.40 & 
    0.99 & 0 & 0 \nl
HD 165401 & -0.44 & 0.90 & 6.80 & 4.86 & 5734 & 4.40 & 4.33 &   41.00 & 
    0.88 & 0 & 1 \nl
HD 166913 & -1.44 & 0.82 & 8.22 & 4.25 & 6175 & 4.25 & 4.61 &   16.09 & 
    1.04 & 0 & 0 \nl
HD 166913 & -1.80 & 0.81 & 8.22 & 4.25 & 6030 & 4.21 & 3.90 &   16.09 & 
    1.04 & 4 & 0 \nl
HD 174912 & -0.46 & 0.91 & 7.15 & 4.76 & 5894 & 4.42 & 4.19 &   33.31 & 
    0.61 & 0 & 1 \nl
HD 181743 & -1.66 & 0.77 & 9.68 & 4.95 & 6080 & 4.47 & 4.86 &   11.31 & 
    1.76 & 0 & 0 \nl
HD 181743 & -2.00 & 0.77 & 9.68 & 4.95 & 5910 & 4.42 & 4.25 &   11.31 & 
    1.76 & 4 & 0 \nl
HD 184499 & -0.53 & 0.90 & 6.62 & 4.10 & 5750 & 4.11 & 3.98 &   31.29 & 
    0.62 & 0 & 0 \nl
HD 187691 & 0.08 & 1.01 & 5.12 & 3.68 & 6074 & 4.09 & 4.06 &   51.57 & 
    0.77 & 3 & 1 & 19510+1025 \nl
HD 188510 & -1.37 & 0.71 & 8.83 & 5.85 & 5628 & 4.64 & 5.16 &   25.32 & 
    1.17 & 0 & 0 \nl
HD 188815 & -0.54 & 0.92 & 7.47 & 3.72 & 6201 & 4.06 & 4.38 &   17.82 & 
    1.05 & 0 & 0 \nl
HD 193901 & -1.00 & 0.76 & 8.65 & 5.45 & 5796 & 4.58 & 4.79 &   22.88 & 
    1.24 & 0 & 0 \nl
HD 194598 & -1.01 & 0.83 & 8.33 & 4.60 & 6047 & 4.36 & 4.65 &   17.94 & 
    1.24 & 0 & 0 \nl
HD 194598 & -1.03 & 0.83 & 8.33 & 4.60 & 6047 & 4.35 & 4.34 &   17.94 & 
    1.24 & 0 & 0 \nl
HD 194598 & -1.11 & 0.82 & 8.33 & 4.60 & 6050 & 4.35 & 4.19 &   17.94 & 
    1.24 & 3 & 0 \nl
HD 194598 & -1.40 & 0.80 & 8.33 & 4.60 & 5920 & 4.31 & 4.10 &   17.94 & 
    1.24 & 4 & 0 \nl
HD 198044 & -0.28 & 1.00 & 7.31 & 3.82 & 6101 & 4.15 & 4.06 &   20.09 & 
    0.94 & 0 & 1 \nl
HD 199289 & -0.99 & 0.82 & 8.28 & 4.67 & 5936 & 4.35 & 4.71 &   18.94 & 
    1.03 & 0 & 0 \nl
HD 201099 & -0.47 & 0.91 & 7.59 & 4.10 & 5898 & 4.16 & 4.24 &   20.09 & 
    1.02 & 0 & 0 & 21077-0534 \nl
HD 201891 & -0.94 & 0.83 & 7.37 & 4.63 & 5974 & 4.35 & 4.97 &   28.26 & 
    1.01 & 0 & 0 \nl
HD 201891 & -1.05 & 0.82 & 7.37 & 4.63 & 5948 & 4.34 & 4.14 &   28.26 & 
    1.01 & 3 & 0 \nl
HD 205294 & -0.36 & 1.04 & 6.86 & 3.23 & 6234 & 3.89 & 4.09 &   18.78 & 
    0.86 & 0 & 1 \nl
HD 208906 & -0.65 & 0.85 & 6.95 & 4.62 & 6072 & 4.38 & 4.47 &   34.12 & 
    0.70 & 0 & 0 & 21587+2949 \nl
HD 210752 & -0.59 & 0.86 & 7.44 & 4.56 & 5958 & 4.34 & 4.59 &   26.57 & 
    0.85 & 0 & 0 \nl
HD 215257 & -0.62 & 0.88 & 7.41 & 4.28 & 6002 & 4.25 & 4.45 &   23.66 & 
    0.97 & 0 & 0 \nl
HD 218504 & -0.58 & 0.89 & 8.11 & 4.16 & 5971 & 4.19 & 4.37 &   16.20 & 
    1.01 & 0 & 0 \nl
HD 221830 & -0.45 & 0.95 & 6.86 & 4.31 & 5719 & 4.21 & 4.13 &   30.93 & 
    0.73 & 0 & 1 \nl
BD+21 0607 & -1.57 & 0.79 & 9.23 & 4.77 & 6139 & 4.43 & 4.38 &   12.84 & 
    1.33 & 0 & 0 \nl
BD+26 2606 & -2.29 & 0.77 & 9.73 & 4.79 & 6146 & 4.43 & 4.23 &   10.28 & 
    1.42 & 0 & 1 \nl
BD+29 2091 & -1.89 & 0.73 & 10.26 & 5.38 & 5921 & 4.57 & 4.72 &   10.55 & 
    1.75 & 0 & 0 \nl
BD+51 1696 & -1.26 & 0.73 & 9.91 & 5.58 & 5708 & 4.58 & 4.87 &   13.61 & 
    1.54 & 0 & 0 \nl
BD+54 1216 & -1.49 & 0.79 & 9.73 & 4.81 & 6116 & 4.43 & 4.62 &   10.36 & 
    1.47 & 0 & 0 \nl
\enddata
\tablenotetext{a}{Reference code; 0: Gratton et al. 1997, 1: Edvardsson 1998, 2: Bonnell \& Bell 1993, 
3: Fuhrmann et al. 1997, 4: Magain 1989, 5: Ryan, Norris \& Beers 1996}
\tablenotetext{b}{Isochrone code; 0: within the metallicity range of Bergbusch \& Vandenberg 1992; 
1: [Fe/H] too low ([Fe/H]= --2.26 is assumed in the grid of Bergbusch \& Vandenberg 1992) or [Fe/H] too high ([Fe/H]=--0.47 is used)}
\tablenotetext{c}{Components of Double and Multiple stars: Dommanget \& Nys 1994}
\end{deluxetable}

\newpage


\begin{deluxetable}{cccc}
\tablecaption{Mean difference and standard deviation 
between  spectroscopic and  trigonometric gravities
\label{table2}}
\tablehead{\colhead{Reference}  & \colhead{$<$[Fe/H]$>$} & 
\colhead{$<\Delta\log g>$}  & \colhead{Number of stars}  }
\startdata 
Edvardsson (1988) & $+0.01 \pm 0.11$ & $+0.19 \pm 0.15$ & 7 \nl
Bonnell \& Bell (1993) & $-0.20 \pm 0.27$ & $+0.13 \pm 0.25$ & 4 \nl
Gratton et al. (1996) & $-0.42 \pm 0.55$ & $+0.11 \pm 0.17$ & 214 \nl
Fuhrmann et al. (1997) & $-0.84 \pm 0.86$ & $-0.25 \pm 0.18$ & 9 \nl
Magain (1989) &  $-2.10 \pm 0.41$ & $-0.39 \pm 0.19$ & 8 \nl
Ryan et al. (1996) & $-2.54$ & $-0.40$  & 1 \nl
\enddata
\end{deluxetable}

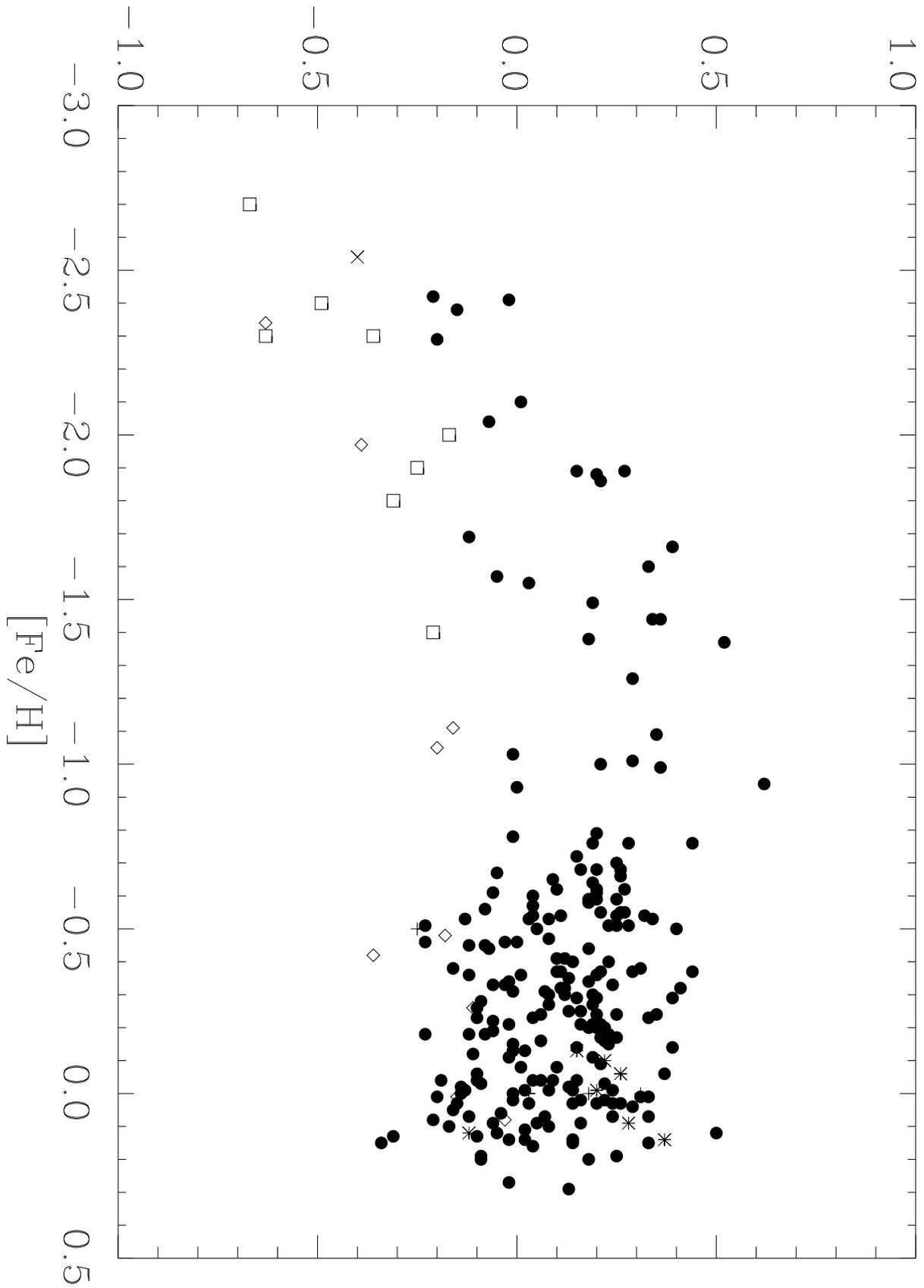
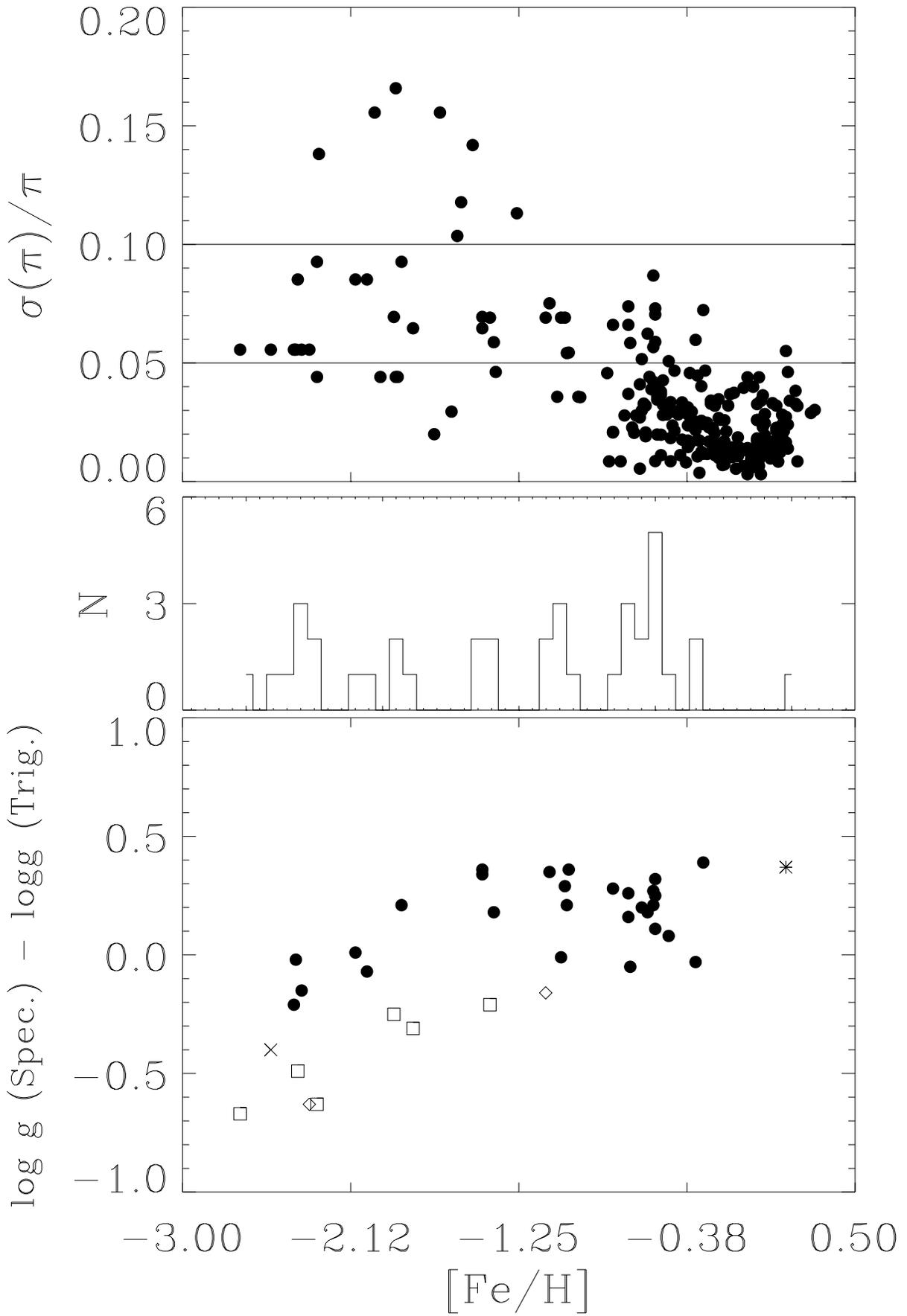
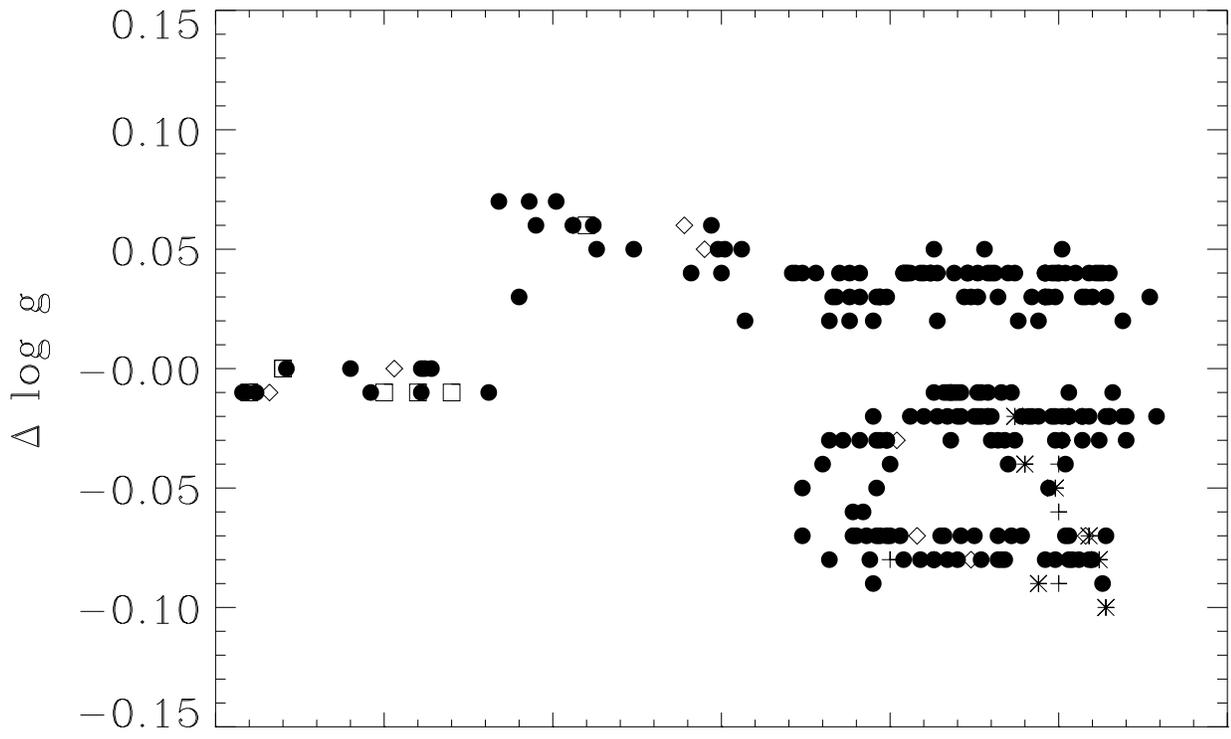
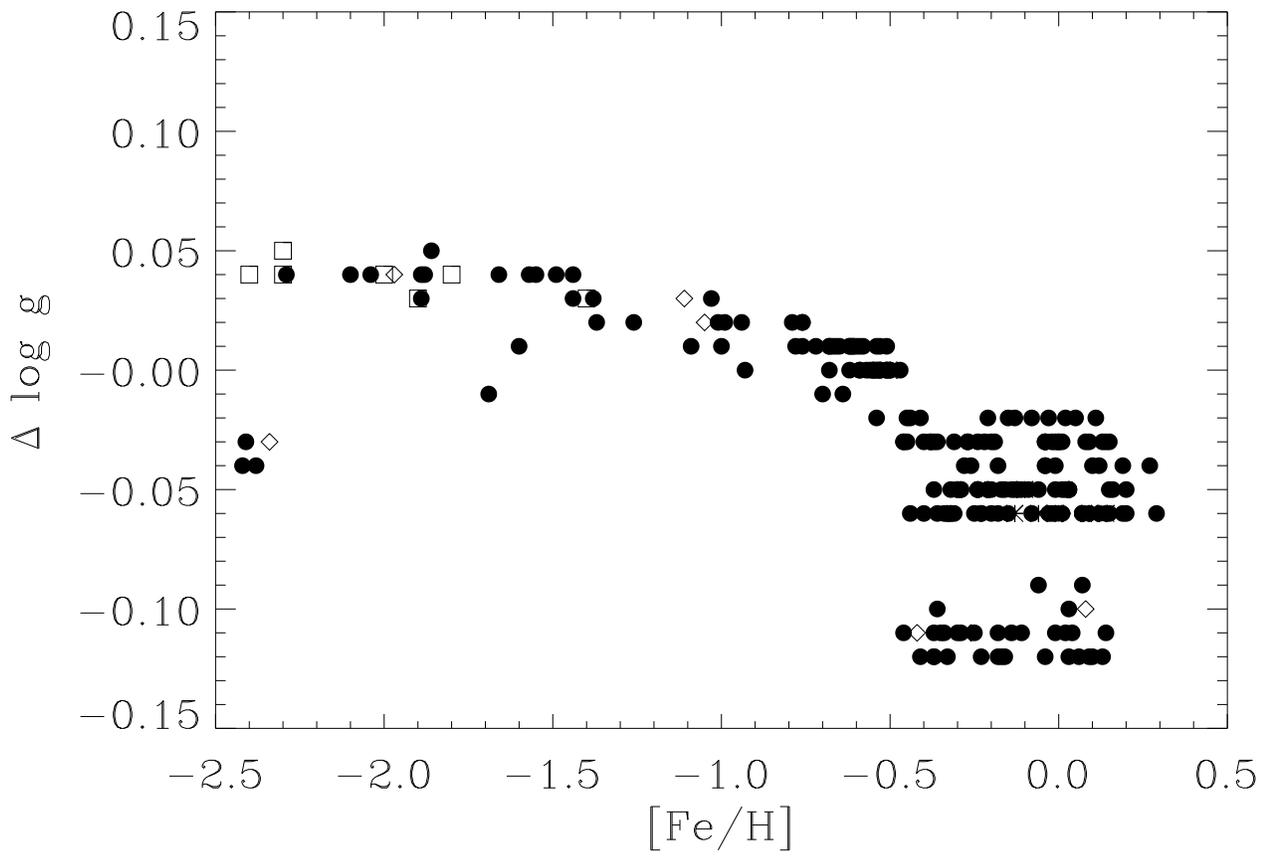
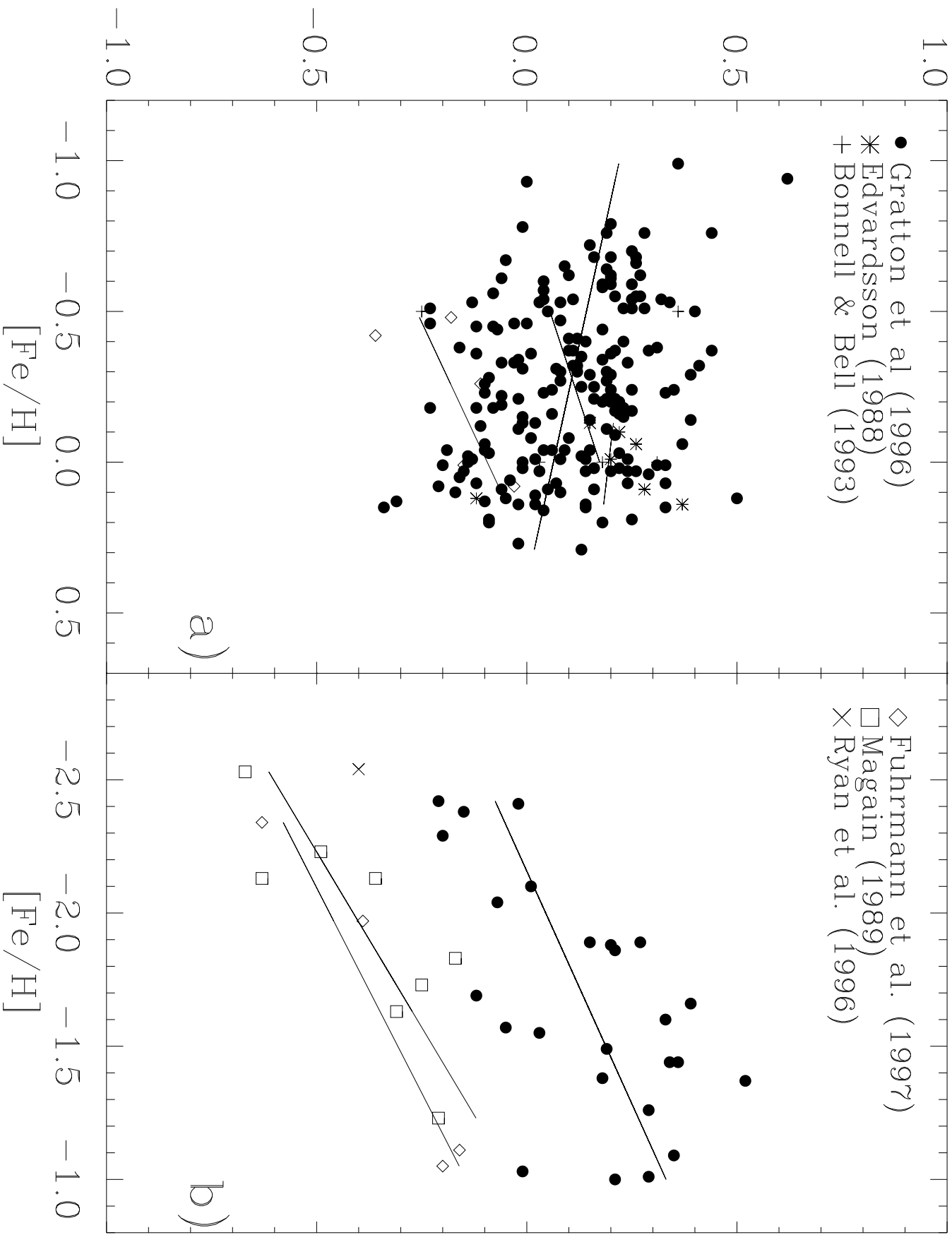
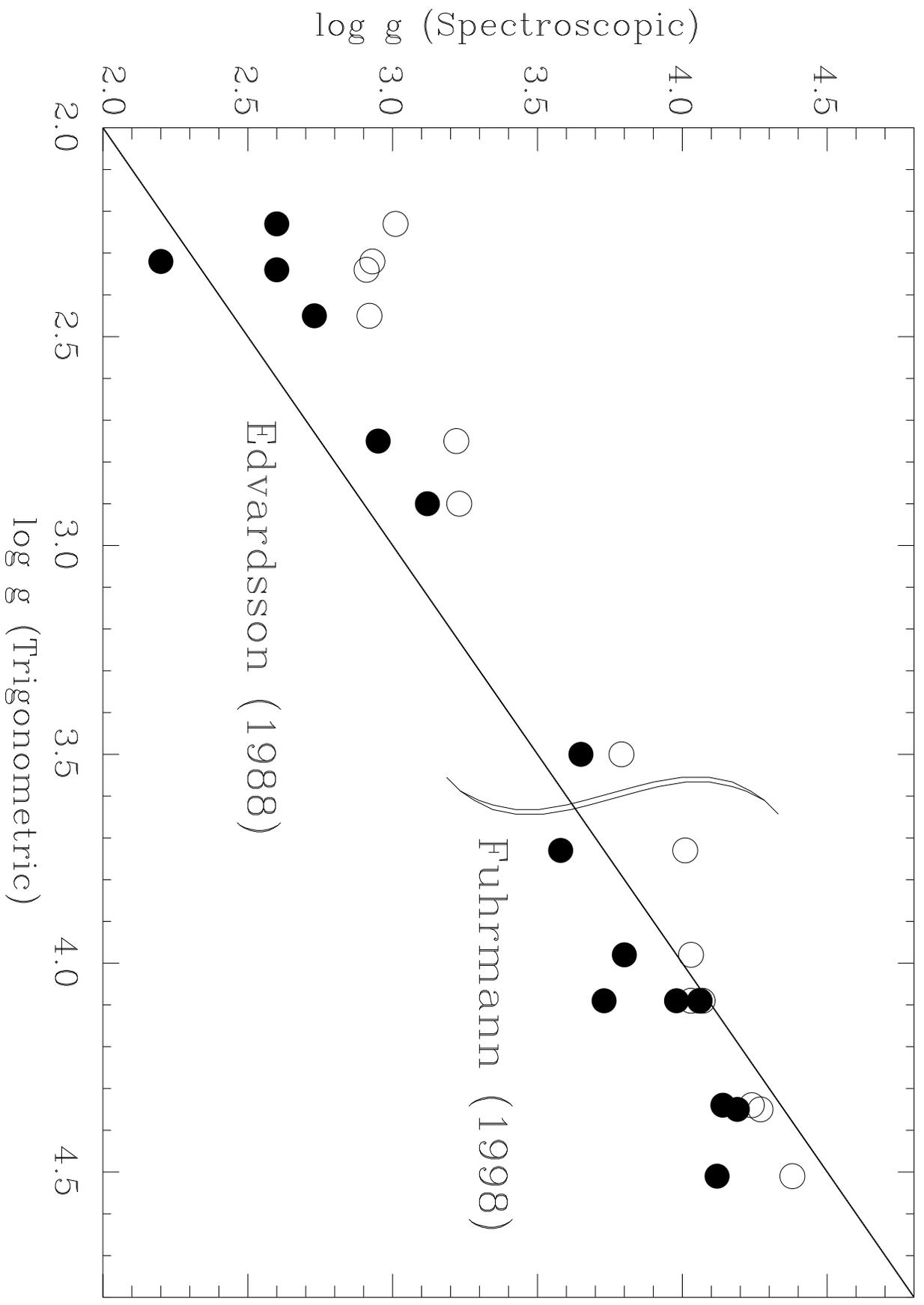
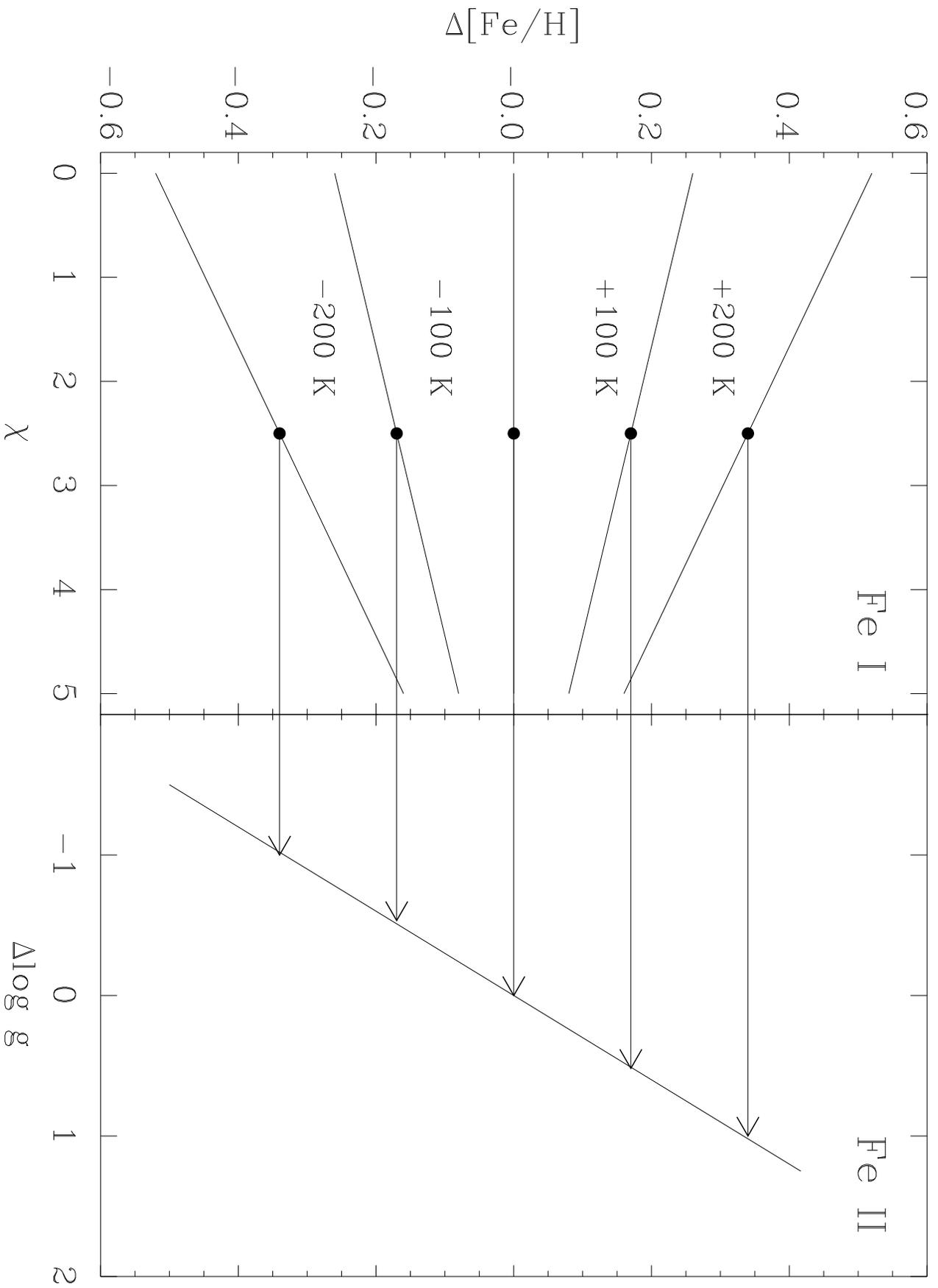
\end{document}